\documentclass{article}
\usepackage{graphicx}
\usepackage[margin=1in]{geometry}
\usepackage{booktabs}
\usepackage{amsmath}
\usepackage{amsfonts}
\usepackage{subfigure}
\usepackage{xcolor}  
\usepackage{authblk}  
\usepackage{cite}
\usepackage{amssymb}
\usepackage{latexsym}
\usepackage{url}
\usepackage{xcolor}
\definecolor{newcolor}{rgb}{.8,.349,.1}

\title{Neural network--based closure models for large--eddy simulations with
explicit filtering}%

\author{Mark Benjamin}
\author{Gianluca Iaccarino}
\affil{Center for Turbulence Research, 481 Panama Mall, Stanford, CA 94305}

\begin{document}
\maketitle
\begin{abstract}
Data from direct numerical simulations of turbulent flows are commonly used to 
train neural network--based models as subgrid closures for large--eddy simulations;
however, models with low \textit{a priori} accuracy have been observed
to fortuitously provide better \textit{a posteriori} results than models with
high \textit{a priori} accuracy. This anomaly can be traced to a dataset shift
in the learning problem, arising from inconsistent filtering in the training
and testing stages.
We propose a resolution to this issue that uses explicit 
filtering of the nonlinear advection term in the large--eddy simulation 
momentum equations, to control aliasing errors. Within the context of explicitly--filtered
large--eddy simulations, we develop neural network--based models
for which \textit{a priori} accuracy is a good predictor of \textit{a posteriori}
performance. We evaluate the proposed method in a large--eddy simulation of a 
turbulent flow in a plane channel at $Re_{\tau} = 180$. Our findings show that
an explicitly--filtered large--eddy simulation with a filter--to--grid ratio
of 2 sufficiently controls the numerical errors so as to allow for accurate
and stable simulations.
\end{abstract}

\section{Motivation}\label{sec:1}
\subsection{The closure problem in LES}
Large--eddy simulations (LES) are becoming commonplace in the numerical
simulation of turbulent flows as a tool for predictive science and engineering.
Advances in
high--performance computing, numerical methods, and computer hardware have 
enabled the use of LES in solving complex engineering problems, such as aircraft
design \cite{kgoc} and natural ventilation in urban landscapes \cite{gorle}. 
The primary modeling challenge in LES is the closure
problem for the residual stresses, and the traditional approach 
relies on using information from the resolved scales. The most 
commonly-used models, such as the dynamic model \cite{dynamic}, leverage scale 
similarity in the inertial subrange \cite{k41}. 
However, the scale similarity assumption
breaks down under a variety of conditions, including ones common
in engineering applications such
as highly anisotropic turbulence, non-equilibrium turbulence, and moderate
Reynolds numbers \cite{sreeni}. 
The limited success of models that use simple, local algebraic representations of 
residual stresses in terms of resolved flow scales has led to an interest in more 
general approaches relying on machine learning (ML)
--- chiefly, artificial neural networks --- to solve the closure 
problem.

\subsection{Neural networks for LES}
The present summer of artificial intelligence \cite{toosi} has seen a 
considerable increase in efforts that employ machine learning techniques to
solve problems in fluid mechanics \cite{brunton}, including turbulence 
modeling \cite{xiao}. Increasing availability of high--fidelity numerical 
simulation data provides fertile ground for using these datasets for
training machine learning models for turbulence. While ML has seen more 
development in Reynolds-averaged
Navier-Stokes (RANS) simulations, its use in LES is rapidly gaining prominence.
In the present work, we focus on machine
learning--based subgrid stress (SGS) models that use neural networks to
learn a functional form for the dependence of the SGS stress (or its divergence,
the subgrid force) on resolved quantities that are computable in the simulation,
such as the resolved strain rate.
This is a supervised learning problem that requires training data with pairs of 
input features
(the aforementioned resolved quantities) and 
the ``true" subgrid stresses in different flow states, which are typically 
obtained by
filtering DNS data.
These models are then applied in the same way
as traditional models: the networks receive resolved quantities as inputs, and 
predict the subgrid stress in each 
grid point and at each time step of the simulation. There have been approaches 
proposed in the literature that leverage differentiable solvers to provide 
ML--based closure models \cite{kochov} \cite{thuerey}; in the present work, we 
limit ourselves to models 
that can be applied in general non--differentiable solvers.

Neural network--based subgrid closures have been studied in
homogeneous, isotropic turbulence \cite{xie} \cite{zhou}, with the conclusion 
that neural network--based models perform just as well or better than
conventional subgrid models in flows they have been trained on. 
However, in applying a similar procedure to the problem of turbulent channel 
flow, Park and Choi \cite{park} observed that models that
had low \textit{a priori} accuracy (measured by correlations of the SGS
stresses predicted by the networks compared to those obtained from the filtered
DNS) showed surprisingly good results in \textit{a posteriori} testing (that is,
in LES that used the neural networks in the solution procedure), while
models with high correlation coefficients (greater than 0.9) were less 
accurate; this finding was also reported by Kang \textit{et al.} \cite{kang}.
These anomalous results are explained by noting that the training procedure 
used for all these works involves an idealized filtering operation
applied to the DNS to assemble the ``true" subgrid stresses, which nonetheless
makes no appearance in the LES that the models are tested
in. Stoffer \textit{et al.} \cite{gmd} accounted for this fact by training a 
model that learnt both the subgrid scales as well as a representation of the 
discretization error in the LES; however, the model was unstable, due to 
compounding of numerical errors in high--wavenumber modes. It has been found
that this dynamical accumulation of errors can be best accounted for by using
a multi--step loss function coupled with a differentiable solver \cite{thuerey}.








While a 
similar anomaly between \textit{a priori} and \textit{a posteriori} performance 
is known to exist in traditional subgrid models \cite{mene},
the nature of the discrepancy is different in the case of neural network-based
models, in that the filtering operation used for the training is different from
the one used in the testing, which is an example of \textit{dataset shift} or
\textit{joint distribution shift} in machine learning. That is, the distributions
of the inputs and outputs to and from the network differ between the 
\textit{a priori} and \textit{a posteriori} stages.
A ramification of this inconsistency is the 
lack of clarity about the performance of a model without expensive 
\textit{a posteriori} testing. 


The objective of the present 
investigation is to
provide a systematic procedure to eliminate this discrepancy, and to develop
neural network-based LES closure models for which the accuracy in the training 
and the inference stage are correlated. We remark here that the terms 
``training stage", ``\textit{a priori} analysis" and ``offline training"
are equivalent and represent the same process; likewise, the terms ``inference
stage", ``\textit{a posteriori} analysis" and ``online training" are 
equivalent. The first pair of terms is more commonly used in the machine
learning community, while the second pair is more popular in the turbulence 
modeling literature. A more detailed description of what both stages represent is 
provided in Section \ref{sec:1}.

The rest of this paper is organized as follows: Section \ref{sec:1} introduces
neural network-based subgrid models and highlights the dataset shift issue in
the context of turbulent channel flow. Section \ref{sec:2} describes the 
classical formulation of explicitly--filtered LES, presents details of the
filtering operators for both homogeneous and inhomogeneous directions, 
and presents a framework for subfilter stress modeling in 
large--eddy simulations using neural networks. Section \ref{sec:3} discusses
the results, both in the offline (\textit{a priori}) mode and in the 
(\textit{a posteriori}) inference LES. 
Finally, Section \ref{sec:4} summarizes the contribution of the present
work, highlights areas that demonstrate promise and present challenges,
and discusses the implications of the developed method.

\section{Neural network-based SGS modeling}\label{sec:1}
\subsection{Implicitly--filtered LES}
The equations for LES of an incompressible, isothermal flow, are 

\begin{equation}\label{eq1}
    \partial_i \bar{u}_i = 0
\end{equation}
and
\begin{equation}\label{eq2}
    \partial_t \bar{u}_i +\partial_j (\overline{u_i u_j}) 
    = \partial_j( \nu \partial_j \bar{u}_i ) - \partial_i \bar{p}/\rho 
\end{equation}
where the unknowns are the filtered velocities $\bar{u_i}$ and the pressure 
$\bar{p}$, and the density $\rho$ and kinematic viscosity $\nu$ are known constants. 
The overbar represents a low-pass filtering operation. The second term on the
left-hand side is the non-linear advection term, which is the only term not
in terms of the primitive variables $\bar{u_i}$ and $\bar{p}$. The following 
decomposition \cite{lenny} is typically employed to rewrite the non-linear 
product:
\begin{equation}\label{eq3}
    \overline{u_i u_j} = \bar{u}_i\bar{u}_j +\tau_{ij},
\end{equation}
where $\tau_{ij}$ represents subfilter stresses that require modeling. This
decomposition gives us an alternative form of Equation \ref{eq2}:
\begin{equation}\label{eq4}
    \partial_t \bar{u}_i +\partial_j (\bar{u}_i \bar{u}_j) 
    = \partial_j( \nu \partial_j \bar{u}_i ) - \partial_i \bar{p}/\rho 
    -\partial_j \tau_{ij},
\end{equation}
in which the advection term is now in terms of the primitive filtered velocities.
The above equation is typically used in practice and referred to as
grid--filtered LES, 
or implicitly--filtered LES, in which the equations are discretized and 
time-marched with no explicit
application of a filter in the solution procedure. 
This is because the 
low-pass effect of common discretization schemes and the grid provides
an implicit filter \cite{rogallo}. 
We remark here that in the literature ILES typically refers to a methodology 
in which 
the discretization operators are
specifically designed
with use as subgrid models in mind \cite{grini}. 
In ILES,
$\tau_{ij}$ assumes the role of the subgrid stress.

\subsection{Training procedure for neural networks}\label{sec:iles_nn}

In the present work, we employ Fully Connected Neural Networks 
(FCNN), a type of artificial neural
network composed of multiple layers of interconnected neurons, or trainable
parameters. Given an
input vector $\boldsymbol{x} \in \mathbb{R}^n$, the FCNN maps it to an
output $\boldsymbol{y} \in \mathbb{R}^m$, through a series of linear
transformations alternating with nonlinear activation functions.
Specifically, for each layer $l$, the intermediate output
$\boldsymbol{a}^{(l)} \in \mathbb{R}^{d_l}$ is computed as:

\[\boldsymbol{a}^{(l)} = \boldsymbol{W}^{(l)}\boldsymbol{a}^{(l-1)} +
\boldsymbol{b}^{(l)},\]

where $\boldsymbol{W}^{(l)} \in \mathbb{R}^{d_l \times d_{l-1}}$ denotes
the weight matrix, $\boldsymbol{b}^{(l)} \in \mathbb{R}^{d_l}$ is the bias
vector, and $d_l$ is the dimension of the layer $l$. The activation function
$\sigma^{(l)}(\cdot)$ is then applied element-wise to the linear output,
yielding the activated output $\boldsymbol{h}^{(l)} = \sigma^{(l)}(\boldsymbol{a}^{(l)})$.
The process is repeated for each layer until the final output
$\boldsymbol{y} = \boldsymbol{h}^{(L)}$ is produced.
The goal of the training is to learn
a relationship between the subgrid stress ($\tau_{ij}$) and the resolved
strain rate ($\bar{S}_{ij}$)
\begin{equation}
    \tau_{ij} \approx F(\overline{S}_{ij}; \Theta),
\end{equation}
where $\Theta$ is the set of learnable parameters in the FCNN.
The training procedure of an FCNN involves optimizing the learnable parameters
$\Theta$ to minimize the discrepancy between the network's output and the target
values. This is achieved using a loss function $L(\tau_{ij}, \hat{\tau}_{ij})$,
which quantifies the error between the predicted output $\hat{\tau}_{ij} =
F(\overline{S}_{ij}; \Theta)$ and the true output $\tau_{ij}$. An optimization
algorithm is employed to iteratively update the parameters based on the gradient
of the loss function with respect to $\Theta$, until convergence is reached. In
the evaluation stage, the network predicts the local--in--time--and--space
subgrid stress, which is added to the LES.
Figure \ref{fig:4} shows a schematic of the entire training and testing process.
The procedure involves two stages (represented in the figure by a dotted box);
the first involves the training of the neural network. DNS snapshots are
filtered (represented in the figure as a convolution operation with a kernel
$G$) to obtain filtered (or resolved) quantities and residual (or subgrid)
quantities, which are used to assemble training pairs for the neural network.
The network is then trained, resulting in a set of model parameters ($\Theta$).
In the second stage, the trained network is integrated into the LES solver, 
supplying the subgrid stress at each grid point and in each time step. The
third column in the figure represents the proposed approach, involving the 
inclusion of the same filter $G$ in the online LES, which will be discussed
in Section \ref{sec:2}.

While the training procedure as described above is in general applicable to 
any type of flow, it is difficult to make \textit{a priori} assessments of 
generalizability to untrained problems. In the present work, we focus on the
problem of developing models that are performant when the flow parameters in
the LES match those of the DNS used for the training set, with the understanding
that methods developed in the literature for generalization, such as ones 
involving anisotropic filtering of the training set \cite{prakash} or 
transfer learning \cite{subel} can be applied on top of the proposed 
framework.

\begin{figure}%
    \centering
    \includegraphics[width=0.9\textwidth]{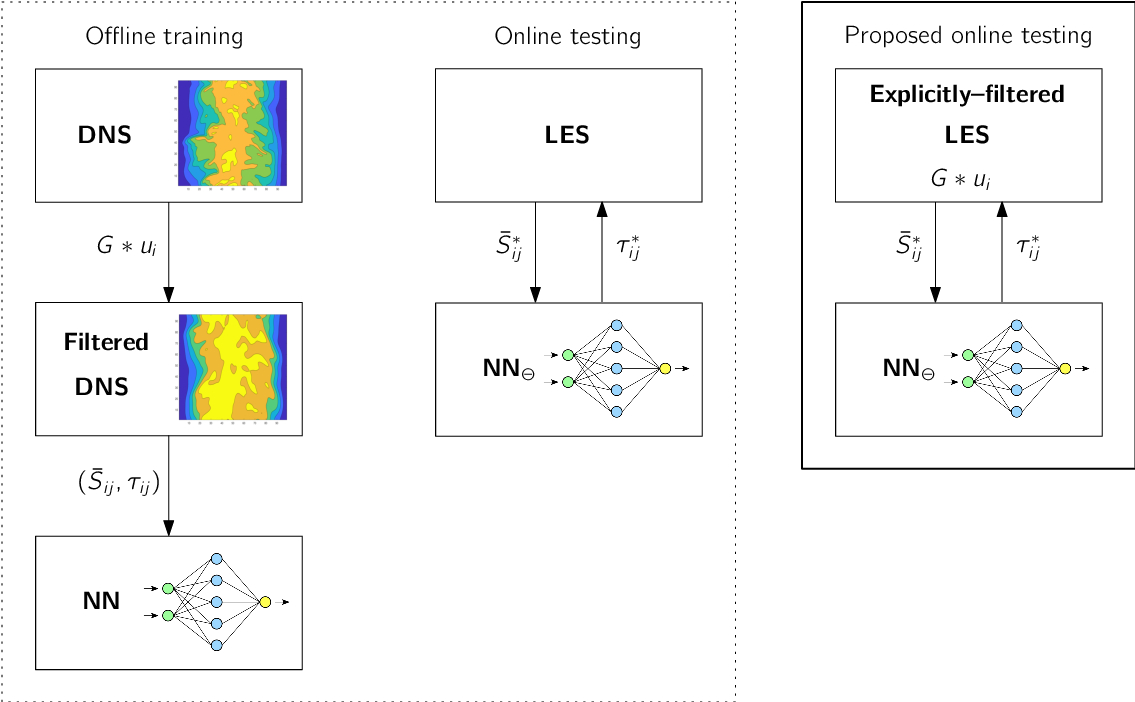} 
    \caption{Schematic of the training and testing phases of a neural network--based
    subgrid model. The filtering operation applied to the training DNS dataset
    is represented as a convolution of the velocity field $u_i$ with a filter 
    kernel $G$. The superscript asterisk is used to distinguish quantities 
computed in the online LES as opposed to obtained from filtered DNS data.}
    \label{fig:4}%
\end{figure}

We consider the turbulent flow in a plane channel at $\textit{Re}_{\tau} = 180$
\cite{moser}.
While this flow is known to have low Reynolds number 
effects, with a log layer not as well defined as in flows at higher Reynolds
numbers, it is nonetheless a useful problem to evaluate a neural network--based method
that does not assume the existence of a log law or use near--wall corrections.
A second-order, staggered finite difference discretization is used, 
which solves the LES equations using a pressure 
projection approach. The convective terms are discretized in the divergence 
form. Time integration is done using a third-order Runge--Kutta 
method, which provides a balance between computational cost, storage and the
time--accuracy required in scale--resolving simulations. 
The code has been validated in prior numerical experiments \cite{stag}.
The flow is driven with a fixed streamwise pressure gradient. The details of 
the grids for the DNS and the LES are provided in Table \ref{tab:22}.

\begin{table}
\centering
\caption{Grid details of channel flow direct numerical simulations (DNS) and
    large--eddy simulations (LES). The grid is uniform in the 
streamwise ($x$) and spanwise ($z$) directions and stretched using a 
hyperbolic tangent function in the wall-normal direction ($y$).}
\label{tab:22}
\begin{tabular}{lllll}
    & $N_x \times N_y \times N_z$ & $\Delta x^+$ & $\Delta y_{min}^+$ & $\Delta z^+$ \\ \hline
DNS & $96 \times 97 \times 96$    & 11.8       & 0.1              & 5.9        \\
LES & $48 \times 48 \times 48$    & 23.6       & 0.2              & 11.8
\end{tabular}
\end{table}

\subsection{Accuracy of NN--based SGS models}
\begin{figure}%
    \centering
    \includegraphics[width=0.9\textwidth]{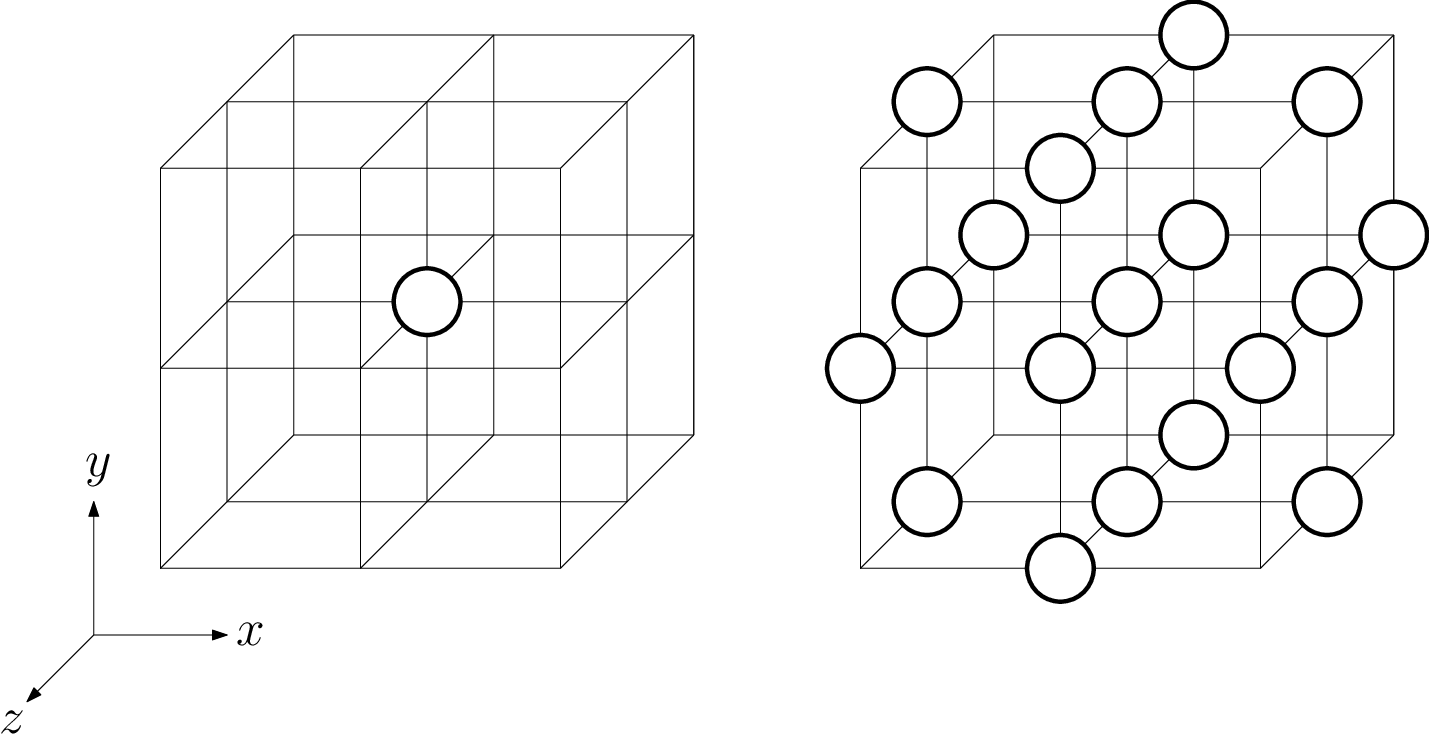} 
    \caption{Schematic of the input stencils for the two neural networks in the
        present study. The circles represent nodes in the structured Cartesian
        grid. Left: 1--point stencil; right: 19--point stencil. The output of 
        the network in both cases is applied to the central node.}
    \label{fig:26}%
\end{figure}

The accuracy of a neural network--based SGS model depends on the model 
architecture and the data used for training. Assuming, however, these are 
optimized by parameter sweeps, there remains the question of poor accuracy 
due to underfitting from choice of input features. Even with the right choice
of network architecture and a substantial and varied dataset, the SGS model
may be unable to learn the the underlying structure of the SGS stress, leading 
to a model with low accuracy. While it is common practice to model the SGS 
stress as a function of the local resolved strain rate, it is known from 
\textit{a priori} testing that the mapping between these variables is not 
perfect \cite{ferz}. In the context of channel flow, it has been observed that
the addition of the resolved rotation rate tensor or the pressure gradient
has no noticeable effect on the model accuracy \cite{park}. The standard 
recourse in the literature has been to use
strain rate information from neighboring nodes \cite{gamahara} \cite{kang}
\cite{park} \cite{gmd} \cite{liu}, giving non--locality to the model. 
Alternatively, one might view the information from the adjacent nodes as 
being equivalent to supplying the network with information about the gradient
of the resolved strain rate. In the
present work, we build on previous studies carried out in channel flow that 
determined the number of neighboring grid points required to maximize accuracy
\cite{mb}. We report results from two models: one that uses local strain rate
information (a low--accuracy, 1--point model), and one that uses strain 
rate information from
all neighboring nodes separated by at most two edges 
(a high--accuracy, 19--point model). A 
schematic of the input stencils for both models is shown in Figure \ref{fig:26}.

\begin{figure}[ht]
    \centering
    \subfigure[]{
        \label{fig:a}
        \includegraphics[width=0.45\textwidth]{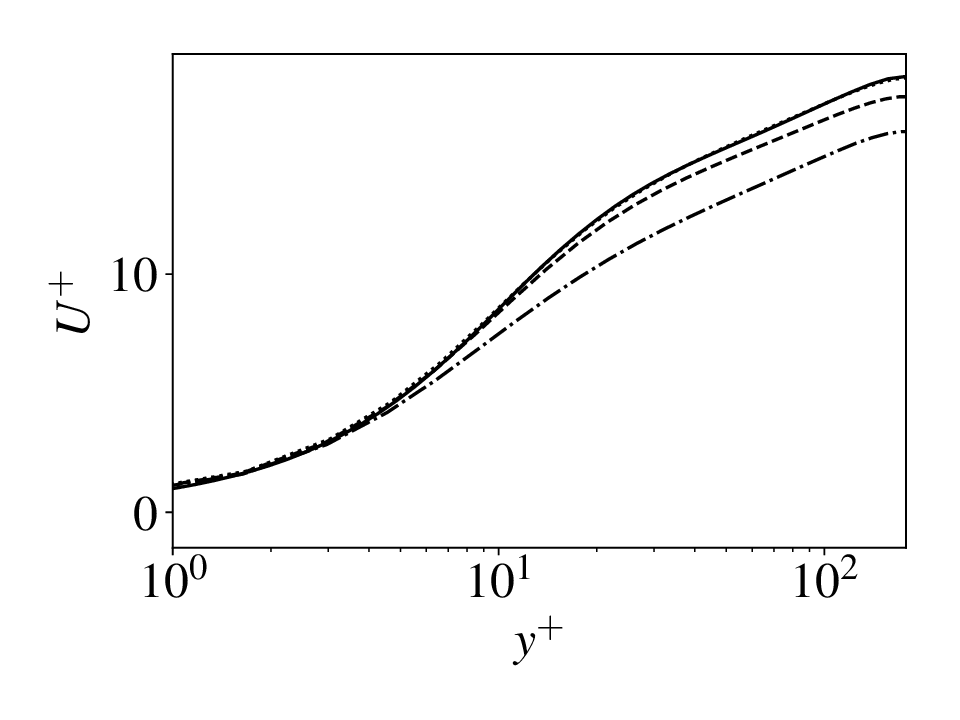}
    }
    \hspace{0.02\textwidth} 
    \subfigure[]{
        \label{fig:b}
        \includegraphics[width=0.45\textwidth]{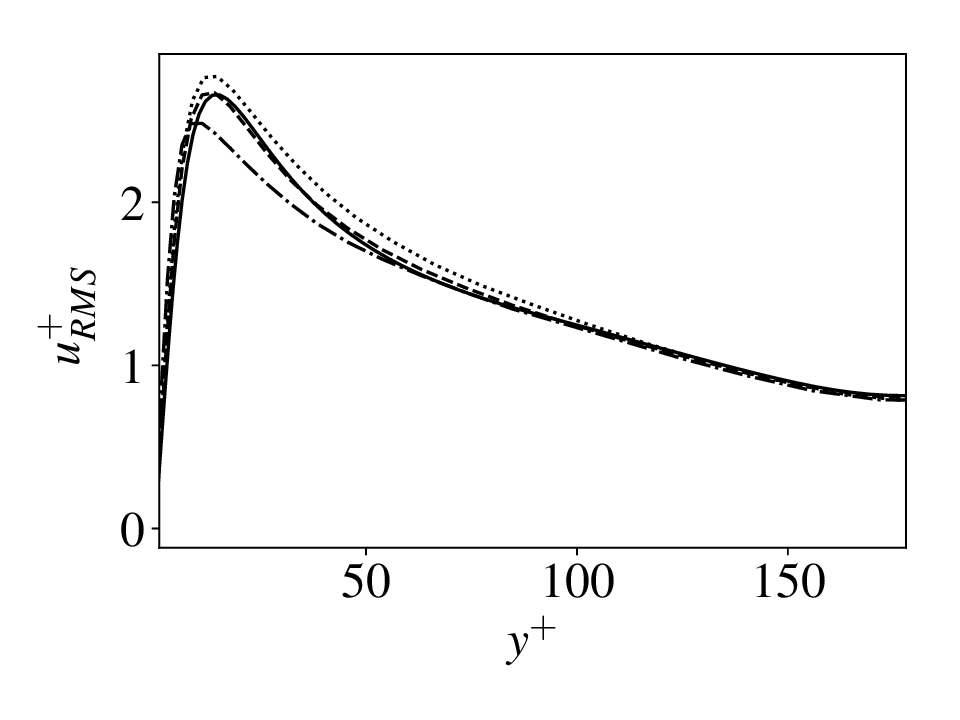}
    } 
    \newline

    \subfigure[]{
        \label{fig:c}
        \includegraphics[width=0.45\textwidth]{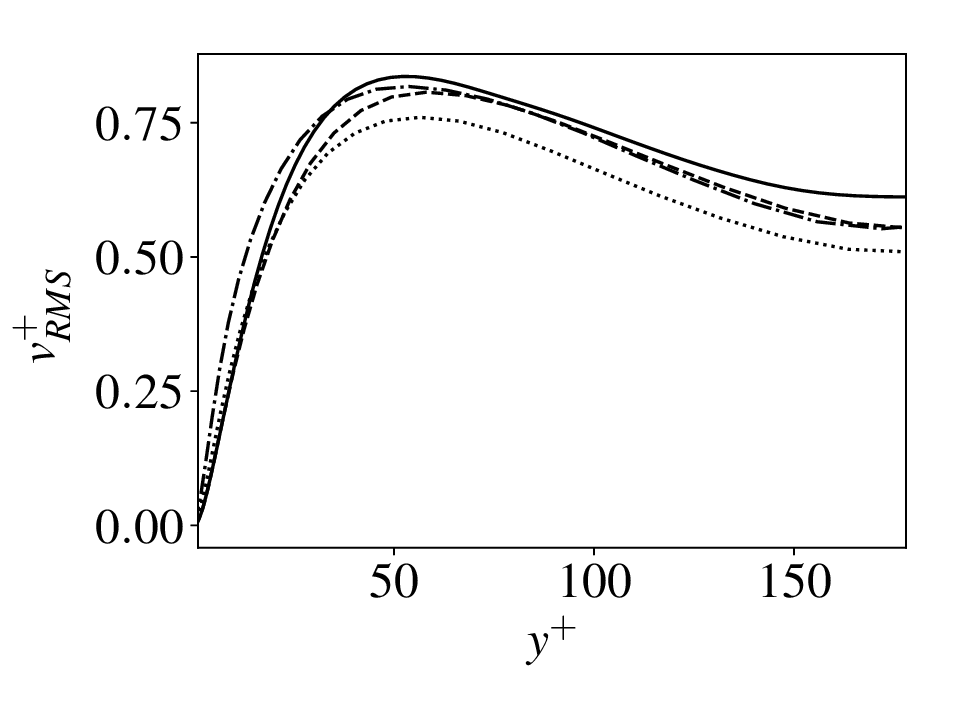}
    }
    \hspace{0.02\textwidth} 
    \subfigure[]{
        \label{fig:d}
        \includegraphics[width=0.45\textwidth]{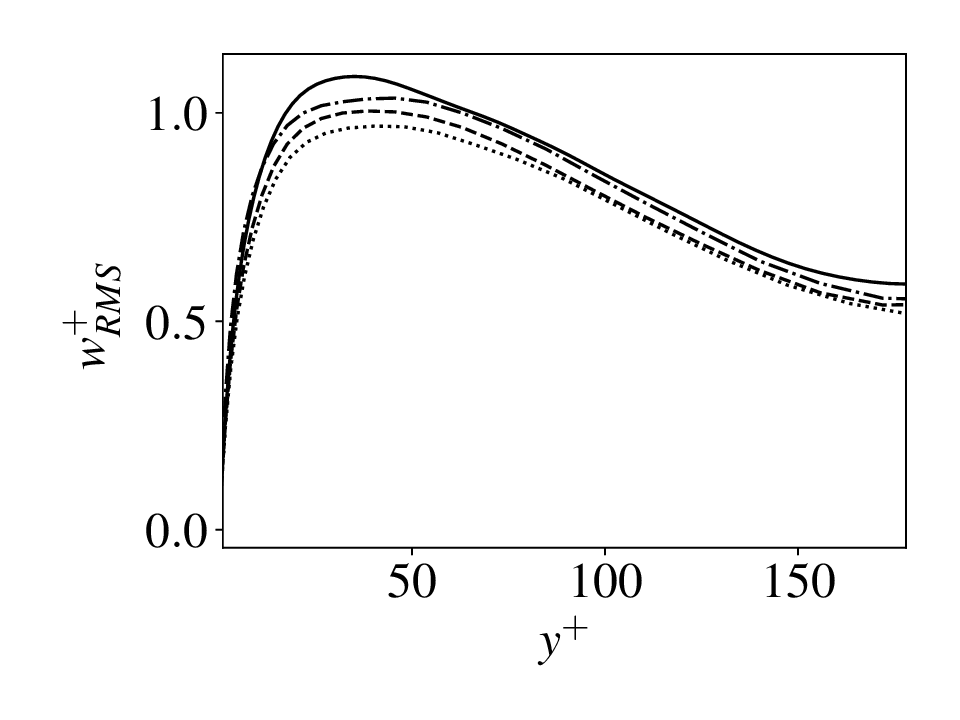}
    }
    \caption{Statistics from ILES: (a) averaged streamwise velocity ($U^{+}$) 
        profiles; (b) streamwise velocity fluctuations; (c) wall--normal
        velocity fluctuations; (d) spanwise velocity fluctuations. Solid line: DNS;
    dotted line: single--point NN-based model; dashed line: 19--point NN-based
    model; dot--dashed line: no SGS model.}
    \label{fig:100}
\end{figure}

Figure \ref{fig:100} (a) shows the results from \textit{a posteriori} testing of both
models \cite{mb}, and shows that the 1--point model predicts the averaged streamwise 
velocity much better than the 19--point model, which actually performs worse
than a simulation that uses no subgrid model at all. As alluded to in Section
\ref{sec:1}, the discretization error overwhelms the modeling error in ILES at
this resolution, and as the networks have not been trained to account for this
error, the performance in the inference ILES bears no relation to the accuracy
\textit{a priori}. That the good agreement of the 1--point model with the DNS
is fortuitous was shown by Benjamin \textit{et al.} \cite{mb} 
by a grid refinement exercise that offset
the balancing effect of the discretization error, revealing the ``true" poor 
performance of the model. The velocity fluctuations, shown in Figure \ref{fig:100}
(b-d) further support the theory that the match is fortuitous, as the 19--point
model is more accurate than the 1--point model in this statistic.

It is clear that models that are agnostic to the 
discretization scheme in ILES will inevitably suffer from poor predictions due
to dataset shift at inference. Two courses of action are possible: the first 
is to approximate the discretization error in the training data, so that the
model learns to account for both modeling and numerical errors simultaneously.
This approach has been tried in the literature and found to lead to divergence
of the solution \cite{gmd} \cite{mb}; the primary issue therein being that the
form of the numerical error poses a harder learning problem than just the SGS
stresses. Moreover, adding unbounded forcing terms to the momentum equations
has implications for numerical stability of the calculation. Finally, this 
results in an intimate coupling between the network and the discretization
scheme used for the inference LES, which prevents the model from being used
with any other discretization. A network, for instance, trained on the error
in a second--order finite difference scheme for the advection term would need
retraining before it can be used with a fourth--order finite difference scheme.
The alternative
is what we propose to do in the present work: to reduce or eliminate completely
the numerical error in the solution by explicitly applying the same filter to
the inference LES that was used to filter the DNS.

\section{Explicitly--filtered LES}\label{sec:2}
The use of an implied filter gives rise to two issues. Firstly, 
finite difference (or finite volume) operators do not provide a purely low-pass
filtering effect \cite{lund}. The second is that linking the filtering operation
to the grid eliminates the notion of obtaining a grid--converged
LES solution. In theory, for a given set of boundary conditions, Equation \ref{eq4} has 
a solution that only depends on the choice of
filter and subfilter model, when numerical errors have been removed by grid
refinement. However, in ILES, grid refinement leads simply to
the exposure of smaller scales, a process that only terminates when the 
Kolmogorov scale for the flow has been resolved. In other words, grid 
refinement of ILES leads to DNS. The issue with this is that all practical ILES
are contaminated with numerical errors, in addition to the modeling errors from
the subgrid stress term, and the commingling of both precludes analysis of
models independent from the numerical scheme used for the solution procedure.

An alternative way of solving the LES equations is to apply an explict filtering
operation to the velocities in the simulation. Such simulations (called ``explicit"
LES, henceforth ELES) typically use the following alternate form of Equation
\ref{eq2}:
\begin{equation}\label{eq5}
    \partial_t \bar{u}_i +\partial_j (\overline{\bar{u}_i \bar{u}_j}) 
    = \partial_j( \nu \partial_j \bar{u}_i ) - \partial_i \bar{p}/\rho 
    -\partial_j \tau_{ij},
\end{equation}
which ensures that the frequency content of all the terms in the equation are 
controlled by the filter, in particular, the advection term.

Although this ELES formulation was advocated already decades ago \cite{moin}, 
it has not been commonly used because it was observed
that the cost of obtaining a grid-independent ELES was comparable to
the cost of a DNS \cite{lund}. Moreover, the subfilter models used in ELES 
performed
poorly in comparison with the subgrid models in ILES, even though the two
are conceptually similar. There are two reasons for this: the first is that 
modeling the subfilter scales poses a stiffer challenge than just the 
subgrid scales, since the filters are larger than the grid, leading to fewer
resolved scales and more scales that require modeling. The second reason is 
that the errors introduced by most numerical schemes used
for ILES have a countervailing effect on the modeling error, leading to results
in a posteriori simulations that are in general better than one might
expect from a priori estimates \cite{ghosal}. At present, explicit 
filtering is primarily used in the context of approximate deconvolution 
methods for LES,
which attempt to reconstruct the unfiltered velocities from the filtered 
velocities \cite{stolz}, and in the determination of model coefficients in 
subgrid models that use the dynamic procedure \cite{dynamic}.

\subsection{Filtering}

For evident reasons, the filter used in LES must attenuate high-frequency
signals. A further requirement is that the filter should commute with 
differentiation, which is an assumption used in arriving at Equations \ref{eq1}
and \ref{eq2}. 
While this is automatically true of a spatially uniform
filter, turbulent flows of interest have directions of inhomogeneity (for
instance, due to the presence of a wall), which necessitate the usage of
a filter of variable
size in order to capture the energy-containing eddies --- the premise of 
large--eddy simulation. To this end, discretely commutative filters have been 
developed, notably those by Vasilyev \textit{et al.} \cite{vasy} for non-uniform 
meshes, with
extensions to unstructured meshes \cite{mars}.

The methodology developed by Vasilyev allows for construction of filters with
a commutation error that scales with an arbitrary power of the mesh spacing.
Thus, by choosing this power to be equal to or greater than the order of accuracy
of the numerical method, the commutation error can be controlled. The 
filtered velocity in the $j$th node of a structured, Cartesian grid is obtained 
as follows:
\begin{equation}
    \bar{u}_j = \sum_{l = -K_j}^{L_j} w_l^j u_{j+l},
\end{equation}
where $w_l^j$ is the filter weight at the $l$th location, and $K$ and $L$ form the
support of the filter. A three-dimensional filtering is obtained by simply 
applying this one-dimensional filter recursively in the three directions.

\begin{figure}[ht]
    \centering
    \subfigure[]{
        \label{fig:a}
        \includegraphics[width=0.45\textwidth]{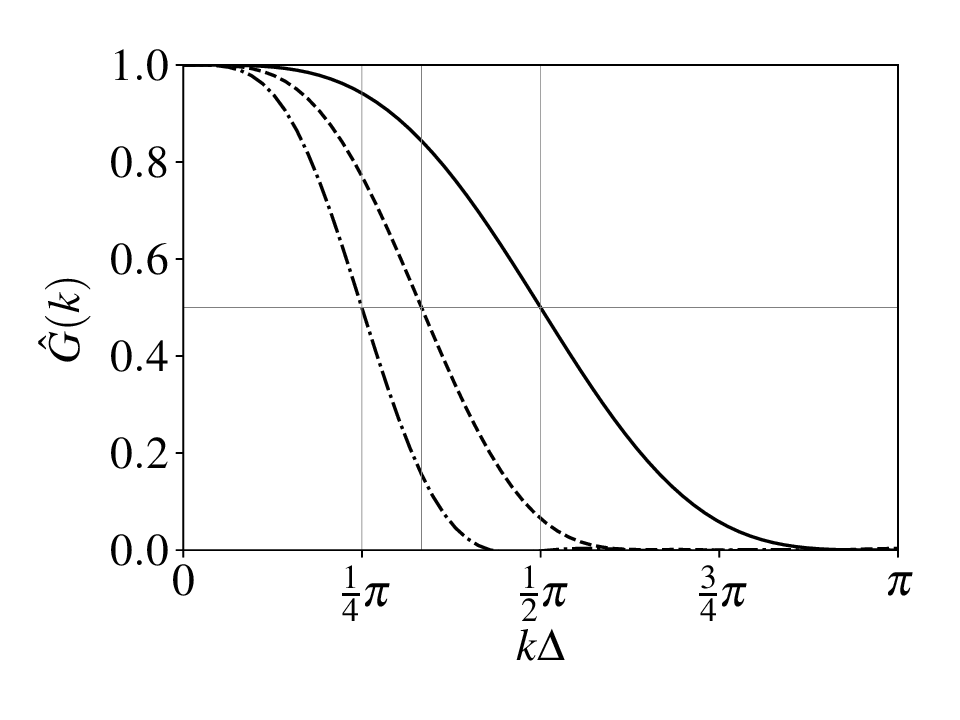}
    }
    \hspace{0.02\textwidth} 
    \subfigure[]{
        \label{fig:b}
        \includegraphics[width=0.45\textwidth]{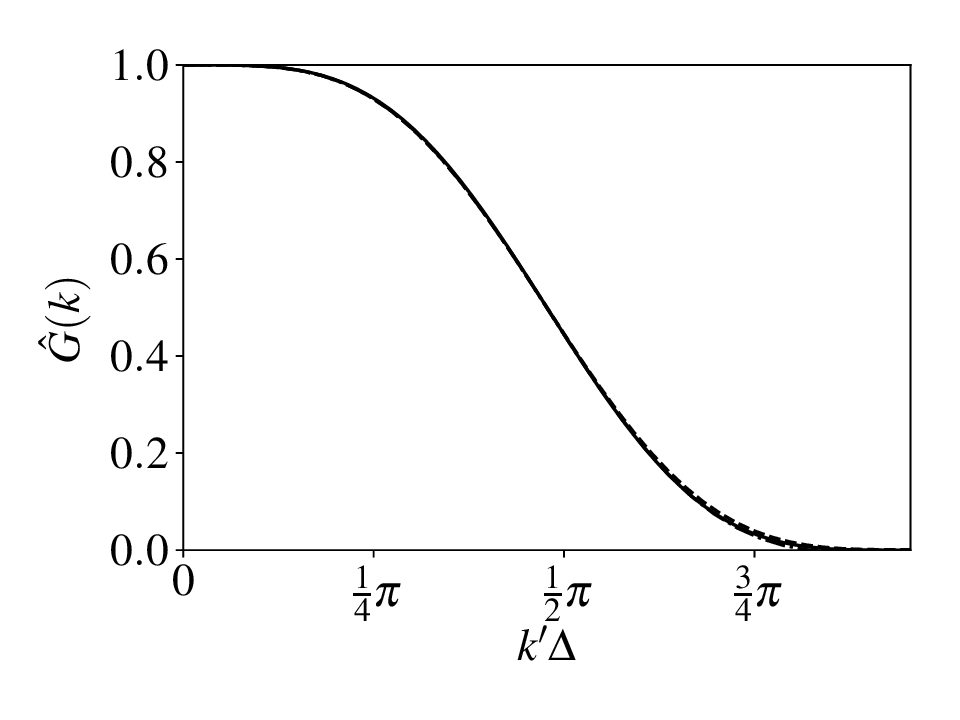}
    }
    \caption{Transfer functions for symmetric, discretely commutative filters
    of different widths from \cite{bose}. Solid line: $2\Delta$; dashed line:
    $3\Delta$; dot-dashed line: $4\Delta$. (a) Transfer functions
    versus nominal grid size $\Delta$. Cross-hairs indicate nominal filter width.
    (b) Transfer functions versus scaled grid sizes.}
    \label{fig:1}
\end{figure}

The definition of the filter in physical space is used operationally, as the 
LES code solves for the velocities in physical space. However, for 
the filters to consistently attenuate the same scales across different grid
resolutions, the 
transfer function, that is, the Fourier transform associated with the filter:
\begin{equation}
    \hat{G}(k) = \sum_{l = -K_j}^{L_j} w^j_l e^{-i\Delta k l},
\end{equation}
must be identical for all grids. In addition, while ILES has a single parameter 
($\Delta$) that describes both the 
characteristic grid and filter width, ELES has a grid size
($\Delta$) and a filter size that is typically defined as a multiple $m$ of the 
grid size ($m\Delta$). Because of this, both the filter size and grid size 
must be specified in order to ensure the filter behaves consistently as the
grid is refined. 
One may instead choose to specify the grid size and a
ratio of the filter size to the grid size, called the Filter--Width Ratio (FWR).
Various studies have been carried out that demonstrate the relationship 
between the FWR and the order of accuracy of the numerical method 
\cite{destef}\cite{chow}\cite{vasiada}. In this work, we follow the same design of 
experiments as Bose \textit{et al.} \cite{bose}, where three grid refinement
levels are considered. Figure \ref{fig:1} (a) shows the transfer functions of the
filters used in the three different grids; the transfer
functions are Gaussian, and of good quality, i.e., there are no overshoots
that may spuriously inject energy into the system. We note here that because the
filters are not sharp cutoffs in spectral space, we can not apply the filter to
the velocity field at the end of a time step; rather, we apply it directly to
the non-linear term in the projection step of each iteration, as described in
\cite{lund}. Figure \ref{fig:1} 
(b) shows the same transfer functions plotted against a modified wavenumber that
accounts for the resolution of the grids they are used in, which shows that,
as expected, all filters filter the same physical length scales. The results 
are shown for the coarsest grid, which matches the computational cost of the 
ILES ($48^3$) in terms of degrees of freedom.

\subsubsection{Wall-normal filtering}
The channel flow has two periodic, homogeneous directions in which a
symmetric filter can be applied. The wall-normal direction, however, requires
asymmetric filtering near the wall. These one--sided filters are developed
using the quadratic optimization procedure described by Vasilyev \cite{vasy},
which minimizes the differences between the real and imaginary parts
of the transfers functions of the asymmetric filter and a symmetric filter of 
the same width. This approach works reasonably well for the near--wall points,
save for the first cell off the wall, which requires a purely one--sided
filter. However, for the wall--resolved LES under consideration, this cell
is within the viscous sublayer; thus the broadband spectral content of the 
velocity field is limited and therefore insensitive to the filtering \cite{bosecomm}.
Thus, no filter is applied at the first cell.
Details of the grids used for the ELES are shown in Table \ref{tab:2}.

\begin{table}[]
\centering
\caption{Mesh resolution and filter width details for ELES. The grid is uniform in the
streamwise ($x$) and spanwise ($z$) directions and stretched using a 
hyperbolic tangent function in the wall-normal direction ($y$).}
\label{tab:2}
\begin{tabular}{@{}llll@{}}
\toprule
Refinement level & Grid resolution & Filter width & Effective resolution \\ \midrule
Coarse & $48^3$ & $2\Delta$    & $24^3$               \\
Medium & $72^3$ & $3\Delta$    & $24^3$               \\
Fine  & $96^3$ & $4\Delta$    & $24^3$               \\ \bottomrule
\end{tabular}
\end{table}

\subsubsection{Grid refinement study}
\begin{figure}[ht]
    \centering
    \subfigure[]{
        \label{fig:a}
        \includegraphics[width=0.45\textwidth]{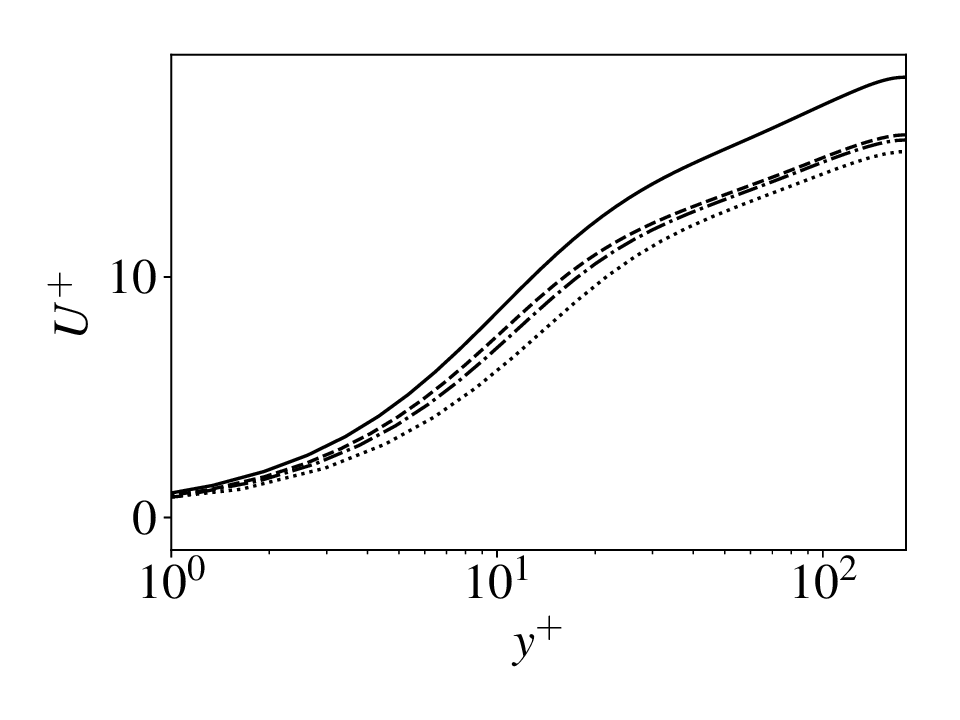}
    }
    \hspace{0.02\textwidth} 
    \subfigure[]{
        \label{fig:b}
        \includegraphics[width=0.45\textwidth]{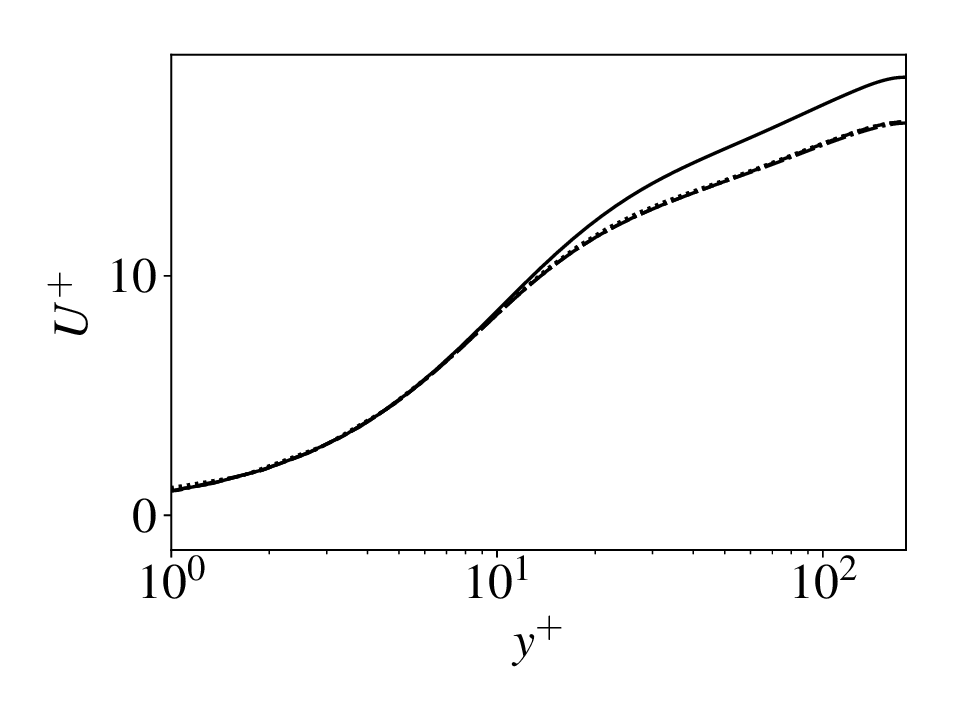}
    } 
    \newline

    \subfigure[]{
        \label{fig:c}
        \includegraphics[width=0.45\textwidth]{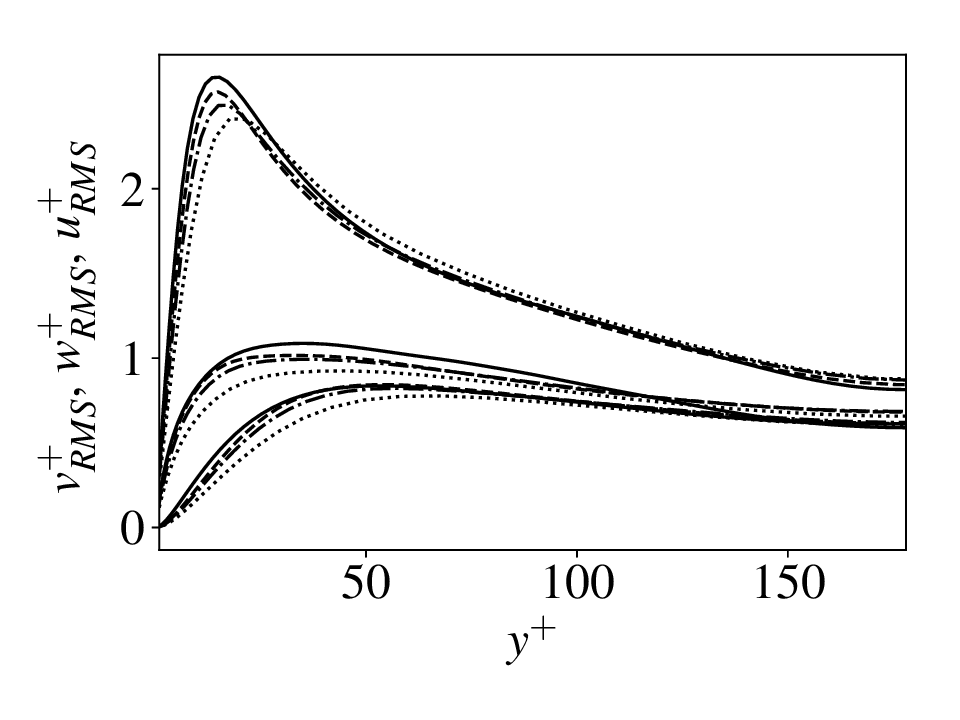}
    }
    \hspace{0.02\textwidth} 
    \subfigure[]{
        \label{fig:d}
        \includegraphics[width=0.45\textwidth]{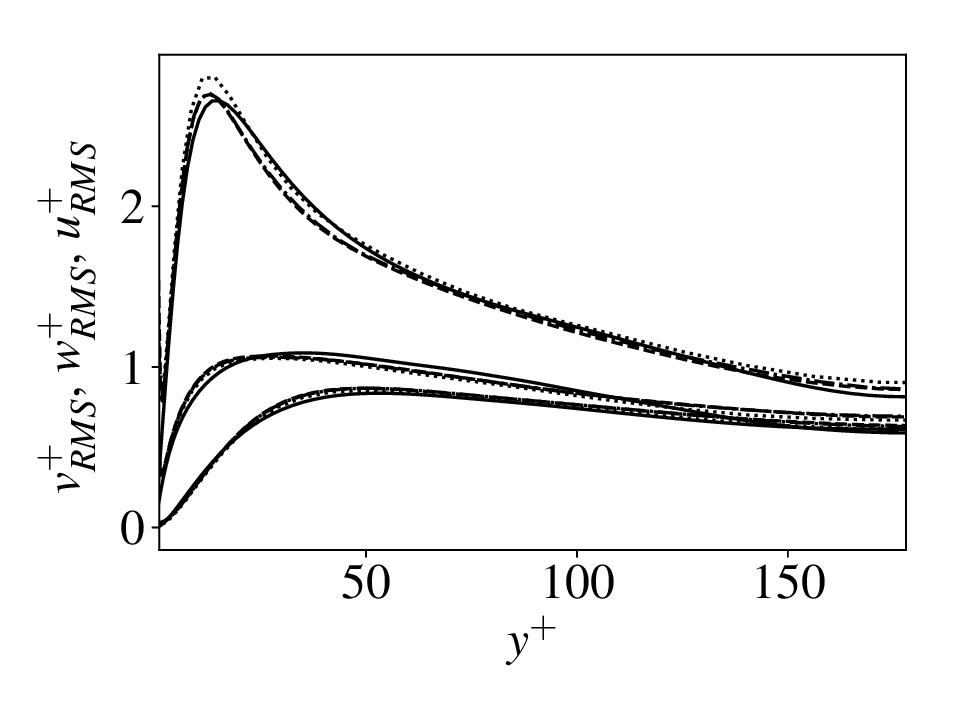}
    }
    \caption{Statistics for ELES: (a) Averaged streamwise velocity ($U^{+}$) with
        static Smagorinsky subfilter model and (b) no subfilter model; (c) 
        Averaged velocity fluctuations with static Smagorinsky subfilter model
        and (d) no subfilter model.
        Solid line: DNS;
        dots: $48^3$ grid; dot--dashed line: $72^3$ grid; dashed line: $96^3$ grid.}
    \label{fig:101}
\end{figure}

We run ELES of the channel at the three grid refinement levels shown above,
first with no subfilter model, then with a Smagorinsky subfilter model with a static
coefficient, which is known to be inaccurate in wall--bounded flows in the
absence of a wall damping function, leading to an over--extended buffer layer, 
but which we nonetheless show here for
demonstrative purposes. Figure \ref{fig:101} (a,b) show the averaged streamwise
velocity profiles for the three refinement levels. We see that while the case
with no SFS model is converged in the mean velocity at all three refinement
levels (the three curves overlap perfectly), while the case with the Smagorinsky 
model is still not fully converged
at the $3 \Delta$ level. The latter behavior is expected; analysis of modeling
and numerical error by Chow and Moin \cite{ghosal}, based on the framework
developed by Ghosal \cite{chowmoin}, recommends a filter width of at least
$4 \Delta$ for grid--independence of ELES with
a second--order finite differences solver. 
The convergence of the case with
no SFS model is 
attributed to the lack of numerical error introduced by the eddy--viscosity closure.
As expected from a simulation with no modeling of the subfilter scales, the 
converged solution is inaccurate. Similar behavior is seen in the higher order
statistics (Figure \ref{fig:101} (c,d)), where the case with the Smagorinsky
subfilter model is unconverged, whereas the case with no subfilter model 
converges at the $3 \Delta$ level.

\subsection{Neural networks for ELES}\label{sec:eles_nn}
\begin{figure}%
    \centering
    \includegraphics[width=0.95\textwidth]{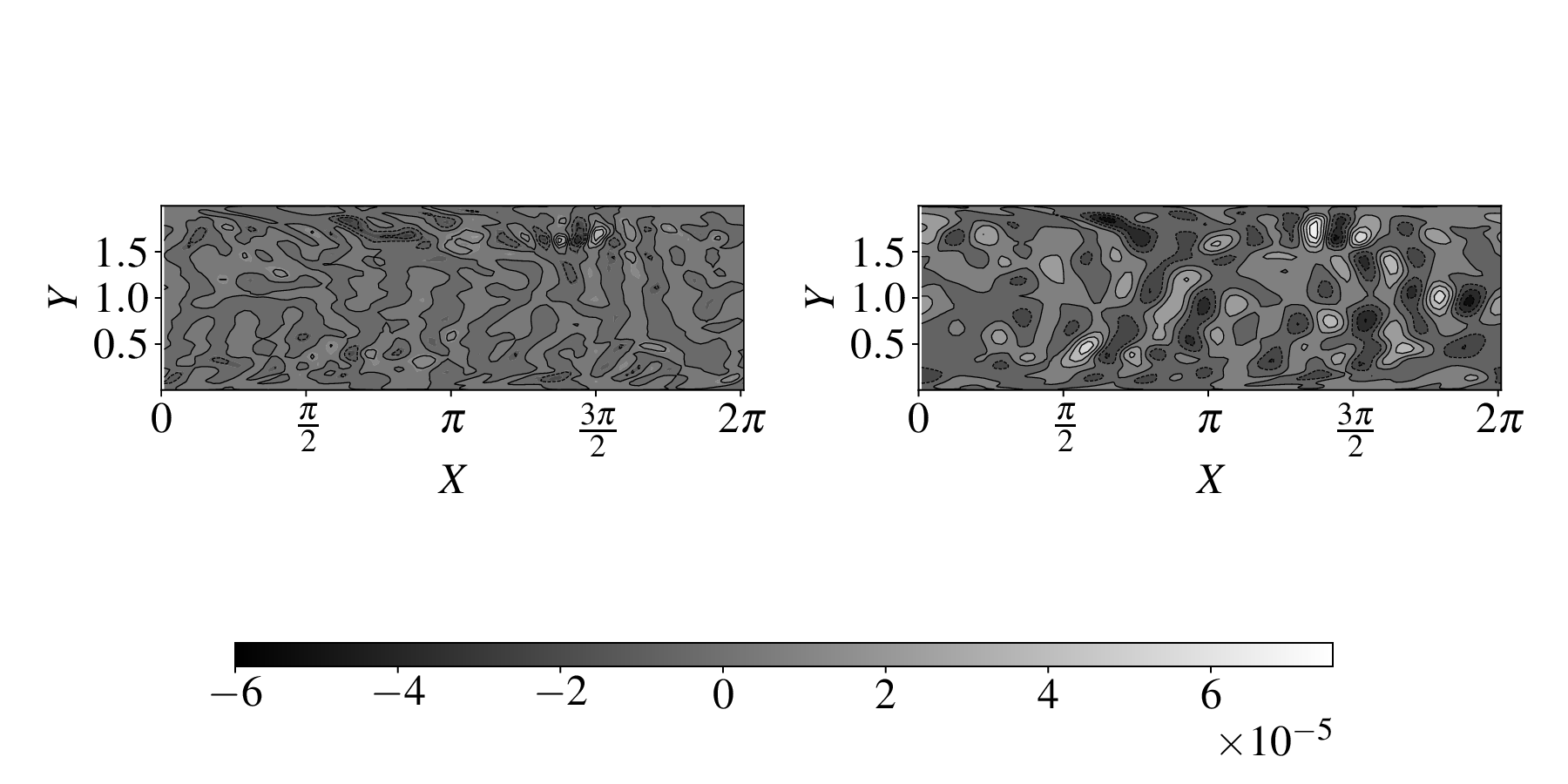} 
    \caption{Contours of $\tau_{12}$ normalized using wall units, from a 
        spanwise plane of filtered DNS, 
        with isocontour lines overlaid.
    Left: filtered with a $2\Delta$ filter for the ILES; right: filtered with 
    a $4\Delta$ filter for the ELES.}
    \label{fig:44}%
\end{figure}

The neural networks for the subfilfter stress in ELES are trained in a manner
identical to that of ILES (see Section \ref{sec:iles_nn}). As with ILES, two
networks are considered: one that uses local strain rate information,
and one that uses a 19--point stencil of strain rates. 
The details of the training procedure are provided in Table \ref{tab:1}.
Since the neural networks will be called at each grid point and each time step,
as described in Section \ref{sec:1}, it
is crucial to maintain a compact architecture to ensure computational efficiency.
The floating-point operations per second (FLOPs) scale with the network's number
of nodes per layer ($N_{nodes}$) and the number of layers ($N_{layers}$), and
can be estimated as $\mathcal{O}(N_{layers} \times N_{nodes}^2)$, and this can
quickly overwhelm the cost of the other terms in the momentum update steps of 
the pressure-projection scheme. For this reason, we avoid the use of dropout
layers \cite{dropout} in the architecture; given the modest size of the networks, it is 
more advisable to simply reduce the size of the network, in case of overfitting.
The networks are initially sized using the work by Park and Choi \cite{park}.
To run the ELES at the same resolution as the ILES ($48^3$ nodes), we apply
a $4\Delta$ filter to the DNS snapshots in the training stage, rather than 
the $2\Delta$ used for the ILES. Thus, the ELES models need to account for 
more scales than the equivalent ILES. This is illustrated in Figure \ref{fig:44},
which shows contours of $\tau_{12}$ at a spanwise plane for the ILES 
and the ELES. The subfilter model for the ELES has to predict stresses
that are not only of a broader range of magnitudes, but also a learn a more 
complex functional form, as there is a natural transfer of information from
the filtered and residual quantities, as the size of the filter increases. 
The limiting size of the filter can be determined from the heuristic that 
large--eddy simulations should resolve about 75\% of the energy in the flow.
For the present work, we merely focus on matching the resolution cost of the 
ILES.

We adopt an eddy--viscosity closure for the subgrid stresses, given by
\begin{equation}
    \tau_{ij}^d = -2\nu_t^{NN} \bar{S}_{ij}
\end{equation}
where $\nu_t^{NN}$ is the eddy--viscosity, and $\tau_{ij}^d$ represents 
the deviatoric part of the SFS stress; the 
trace being absorbed into the pressure field. It is seen that a direct
prediction of the subfilter stress or subfilter force by the neural network
leads to solutions that diverge \cite{osan} \cite{xiec}; the use of an 
eddy--viscosity model allows
for a natural way of incorporating the impact of the subfilter scales on the
resolved scales by modulating the effective viscosity of the flow. 
To prevent destabilization of
the calculations by excessive backscatter, we restrict the eddy--viscosity
to positive values by applying a ReLU activation to the output layer. We note
here that the choice of closure model --- or the decision to predict a 
turbulent viscosity rather than the stresses or their divergence directly ---
is solely to do with the stability of the inference LES, and does not 
affect the arguments about the consistency of the filtering between the 
training and testing
phases.
Since 
enforcing positivity of $\nu_t$
has the effect of overpredicting the subfilter production term
($\mathcal{P}r = -\tau_{ij} \overline{S}_{ij}$),
we rescale the subgrid stresses to maintain the net production in the domain in
each training snapshot, following similar approaches in the literature \cite{akh}
\cite{park};
however, we apply the rescaling at the training stage, rather than the inference
stage, to avoid computing domain--averages in each timestep of the simulation.
Thus, we first eliminate backscatter:
\begin{equation}
    \tau_{ij}^{>} =
    \begin{cases}
    0, & \text{if } \mathcal{P}r < 0 \\
    \tau_{ij}, & \text{otherwise}
    \end{cases}
\end{equation}
then rescale the subfilter stresses as follows:
\begin{equation}
    \tau_{ij}^* = \tau_{ij}^{>} \left(\frac{ \int \! \mathcal{P}r \, \mathrm{d}V}
    { \int \! \mathcal{P}r^{>} \, \mathrm{d}V}\right)
\end{equation}
where $\mathcal{P}r^{>} = -\tau_{ij}^{>} \overline{S}_{ij}$. 

\section{Model testing}\label{sec:3}

\begin{table}[]
\centering
\caption{Training dataset, network architecture and hyperparameter details
for ELES.}
\label{tab:1}
\begin{tabular}{@{}ll@{}}
\toprule
\textbf{Training data}               &                                                         \\ 
Dataset                     & $96^3$ DNS instantaneous flow fields                    \\
No. of samples              & 750 fields; every 4th point in $x$ and $z$; 41M samples \\
Filter width                & $4\Delta$                                               \\
Data non-dimensionalization & Wall units                                              \\ \midrule
\textbf{Network architecture}        &                                                         \\ 
Network type                & Fully--connected, feed--forward neural network          \\
Network size                & 2 hidden layers; 32 nodes per layer                     \\
Network inputs              & Strain rate tensor $\bar{S}_{ij}$ \\
Network output              & Eddy viscosity $\nu_t$ \\
Activation functions        & ReLU                                                    \\ \midrule
\textbf{Optimization details}        &                                                         \\ 
Loss function               & Mean squared error \\
Optimizer                   & Adam                                                    \\
Learning rate               & $10^{-4}$                                               \\
Batch size                  & 2048                                                    \\
No. of epochs               & 250                                                     \\
Input noise                 & Gaussian; 5\% of input magnitude                        \\ \bottomrule
\end{tabular}
\end{table}

\subsection{A priori analysis}
\begin{figure}%
    \centering
    \includegraphics[width=0.95\textwidth]{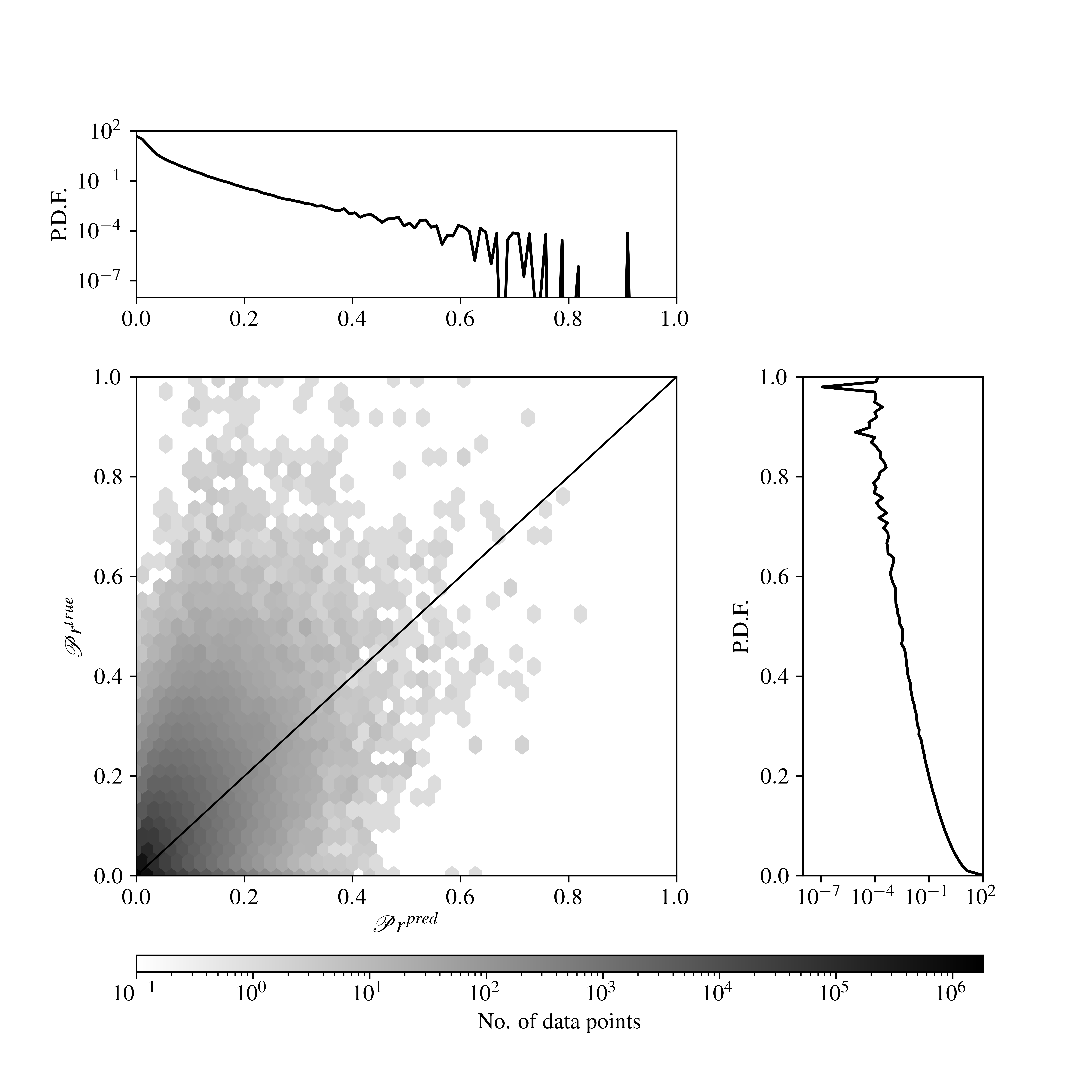} 
    \caption{Correlation plot of subgrid production ($\mathcal{P}r^{pred}$ versus 
        $\mathcal{P}r^{true}$) for the 1--point $\overline{S}_{ij}$ subfilter model,
    with marginal probability density functions (P.D.F.s). Values are normalized
    by wall scaling. The hexagonal bins are shaded
    by density. 4 million data points are represented.}
    \label{fig:3}%
\end{figure}

The accuracy of the two trained neural networks is measured offline by making
predictions on filtered snapshots that were not used for the training. Since
the primary goal of a subfilter model is to dissipate the right amount of 
energy from the resolved scales, we quantify the prediction of the model in
terms of the subfilter production term, which appears in the resolved kinetic
energy equation. For an eddy--viscosity model, this is
given by
\begin{equation}
    \mathcal{P}r = 2\nu_t^{NN} \bar{S}_{ij} \bar{S}_{ij}. 
\end{equation}
Figure \ref{fig:3} shows a correlation plot between the subfilter production
predicted by the 1--point model ($\mathcal{P}r^{pred}$) and the subfilter production obtained by
filtering the DNS ($\mathcal{P}r^{true}$). The Pearson correlation coefficient 
is 0.85. We see a general tendency of the model to underpredict the subfilter
production, from the positive slope of the core of the correlation plot 
relative to the reference line. Additionally, the marginal probability densities 
show that the
tail of the distribution is  poorly predicted. Figure \ref{fig:4} shows the 
correlation plot for the 19--point model; the predictions are much better 
correlated with the truth values, and the tail of the distribution properly
captured, with a correlation coefficient of 0.99. Any scatter observable in
the plot relative to the reference line occurs with low probability, as seen
by the tail of the distribution, which is well--predicted. 

\begin{figure}%
    \centering
    \includegraphics[width=0.95\textwidth]{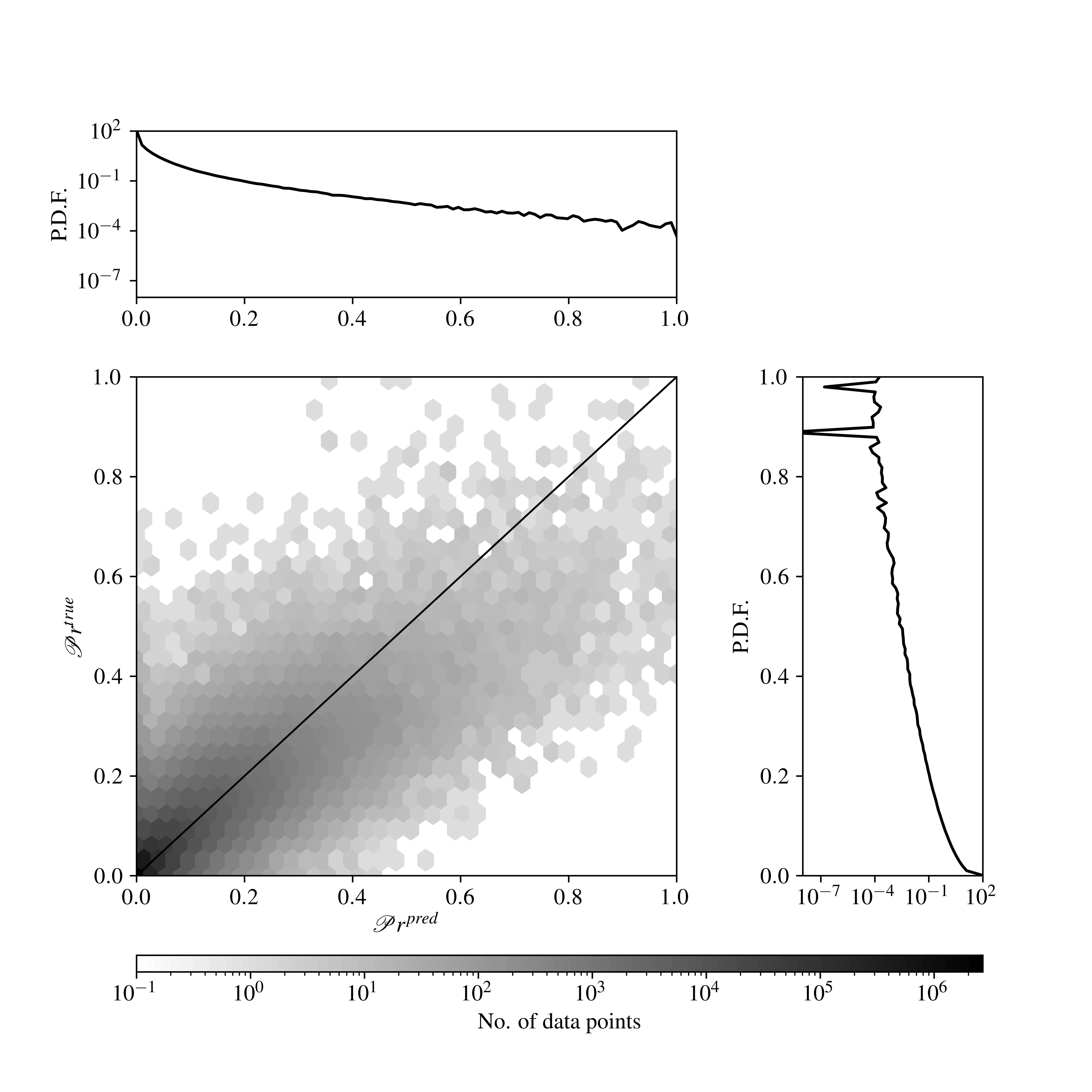} 
    \caption{Correlation plot of subgrid production ($\mathcal{P}r^{pred}$ versus 
        $\mathcal{P}r^{true}$) for the 19--point $\overline{S}_{ij}$ subfilter model,
    with marginal probability density functions (P.D.F.s). Values are normalized
    by wall scaling. The hexagonal bins are shaded
    by density. 4 million data points are represented.}
    \label{fig:4}%
\end{figure}

\subsection{A posteriori analysis}
\begin{figure}[ht]
    \centering
    \subfigure[]{
        \label{fig:a}
        \includegraphics[width=0.45\textwidth]{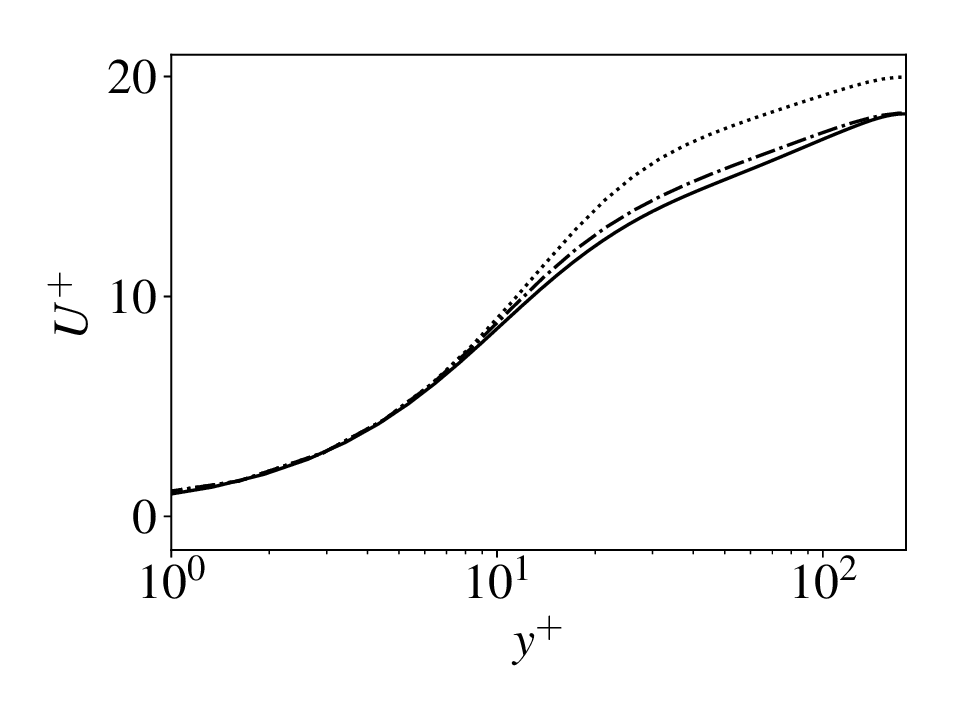}
    }
    \hspace{0.02\textwidth} 
    \subfigure[]{
        \label{fig:b}
        \includegraphics[width=0.45\textwidth]{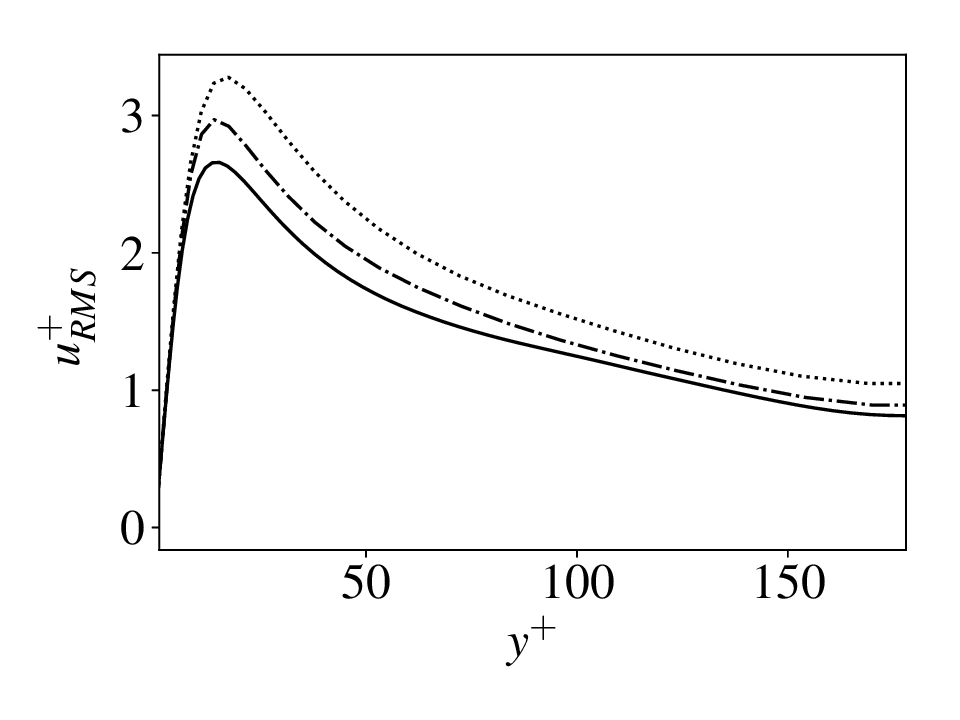}
    } 
    \newline

    \subfigure[]{
        \label{fig:c}
        \includegraphics[width=0.45\textwidth]{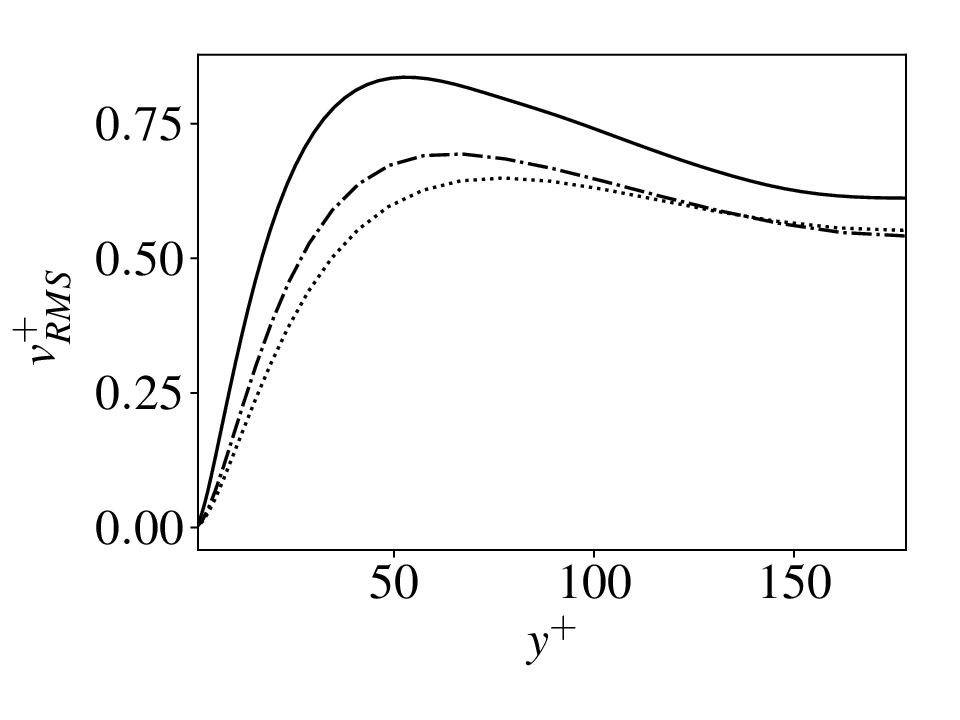}
    }
    \hspace{0.02\textwidth} 
    \subfigure[]{
        \label{fig:d}
        \includegraphics[width=0.45\textwidth]{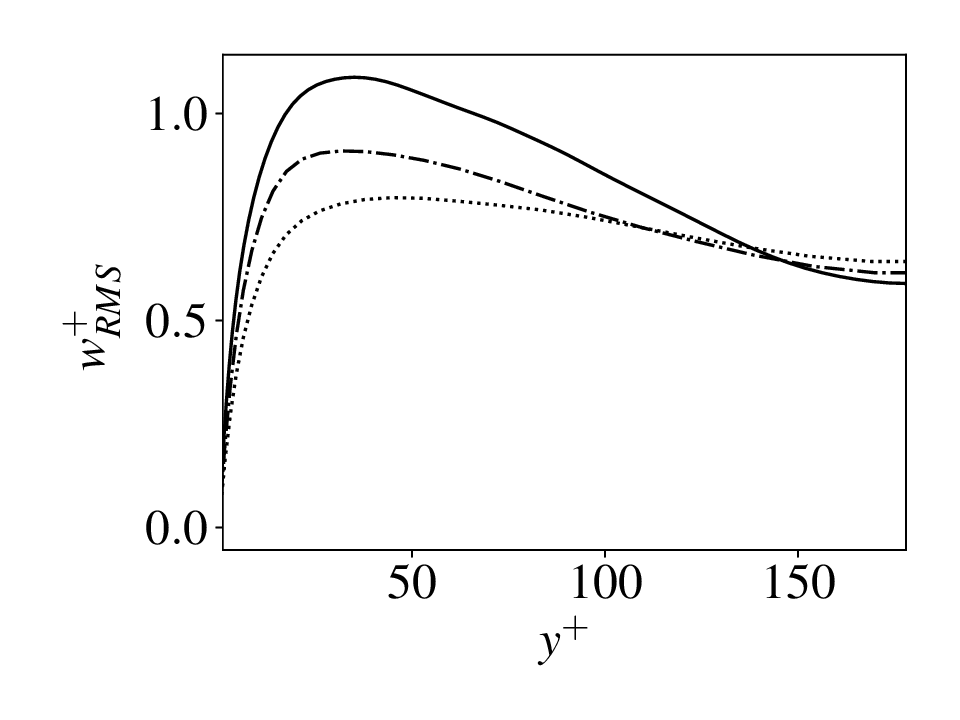}
    }
    \caption{Statistics from ELES: (a) averaged streamwise velocity ($U^{+}$) 
        profiles; (b) streamwise velocity fluctuations; (c) wall--normal
        velocity fluctuations; (d) spanwise velocity fluctuations. Solid line: DNS;
    dotted line: single--point NN-based model; dashed line: 19--point NN-based
    model.}
    \label{fig:102}
\end{figure}

Figure \ref{fig:102} shows the results from the ELES with the two neural network--based
SFS models. We see that the single--point model provides too much dissipation 
and leads to an overprediction of the mass flux. The 19--point model, however,
provides a much more accurate prediction, and we observe a convergence towards
the DNS, as predicted by the \textit{a priori} correlations. Similar convergence
trends are observed in the stresses, with the near--wall peak in the streamwise
fluctuations much better predicted by the 19--point model, though it must be noted that these 
higher order statistics are limited by the linear eddy--viscosity assumption,
which tends to overpredict the streamwise stresses at the expense of the other
two directions. However, the present framework does not place restrictions 
upon the form of the closure; as described earlier, a positive eddy--viscosity
model is used here for stability, but anisotropic models that do not make
assumptions about the alignment between the resolved strain rate tensor and 
the subfilter tensor can be explored in future work to further improve the
accuracy of higher--order statistics.

\begin{figure}[ht]
    \centering
    \subfigure[]{
        \label{fig:a}
        \includegraphics[width=0.45\textwidth]{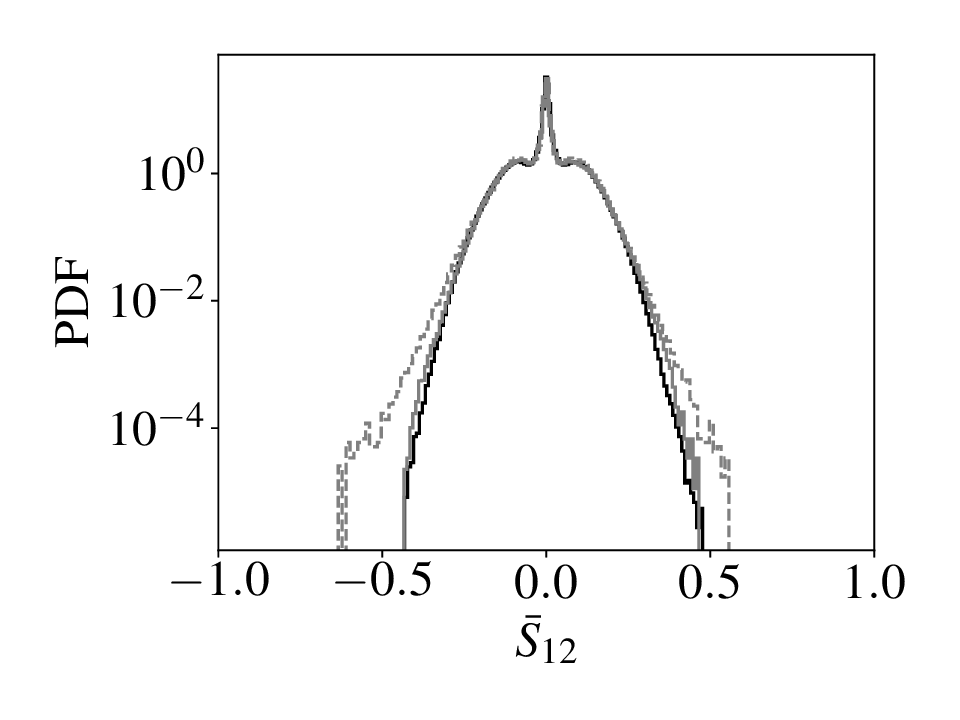}
    }
    \hspace{0.02\textwidth} 
    \subfigure[]{
        \label{fig:b}
        \includegraphics[width=0.45\textwidth]{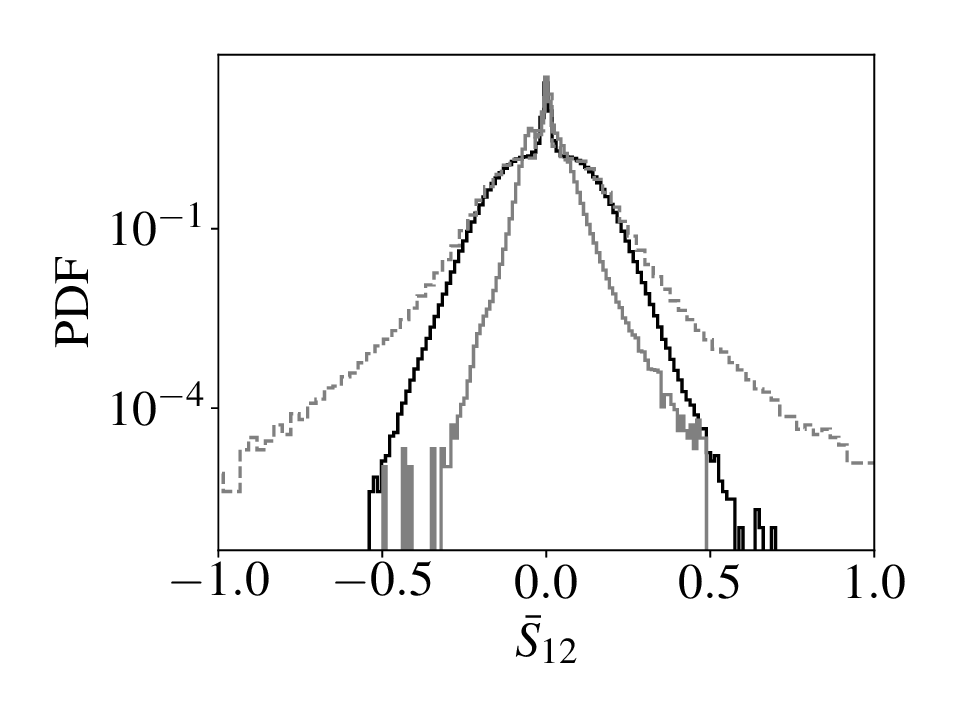}
    }
    \caption{Histograms of $\bar{S}_{12}$ from (a) ELES and (b) ILES. Solid black
        line:
    DNS (filtered at $4\Delta$ for ELES and $2\Delta$ for ILES); dashed gray
    line: 1--pt model; solid gray line: 19--pt model.}
    \label{fig:332}
\end{figure}

To assess the extent to which the ELES approach mitigates the dataset shift,
Figure \ref{fig:332} plots histograms of the $\bar{S}_{12}$ --- the primary
component that is active in a plane channel driven along the $x$-direction ---
and compares those obtained from the filtered DNS used for training with those
obtained from the \textit{a posteriori} LES. We see that when ELES is used, not
only do both the single and multi--point models closely approximate the 
distribution of the DNS, but we again observe convergence with increasing 
model accuracy; i.e., the 19--point histogram is almost exactly aligned with
the DNS. On the other hand, the histograms for ILES clearly demonstrate that
the inference simulation does not ``see" strain rates drawn from a similar 
distribution to the ones used for training the models, for both the 1--point
and 19--point cases. 

\begin{figure}%
    \centering
    \includegraphics[width=0.65\textwidth]{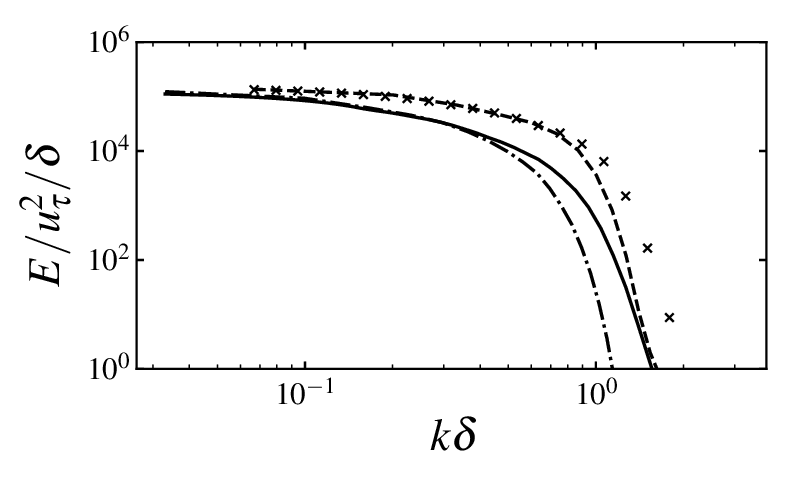}
            \caption{Streamwise velocity spectrum at $y^+ \approx 180$, 
                non-dimensionalized by friction velocity ($u_{\tau}$) and
                channel half-height ($\delta$). Solid line: DNS; dot-dashed 
            lines: DNS filtered at $4\Delta$; dashed line: ELES with 19--pt model; crosses:
            ILES with 1--pt model}
    \label{fig:53}%
\end{figure}

In order to study the effect of the explicit filtering on the spectral content
of the simulation, we show the streamwise energy spectra at the channel midplane
for different simulations in Figure \ref{fig:53}, comparing DNS with filtered
DNS, the single--point subgrid model from ILES (which is the most accurate
in the \textit{a posteriori} test) and the 19--point subfilter
model from ELES. We observe that the ELES shows range of scales that is more 
limited than ILES, and is much closer to the DNS in the tail of the spectrum.
The energy in the large scales is unmodified, as is consistent with our 
expectations. 
Finally, we note that while the velocity profiles for the single--point subgrid
model in ILES matched the DNS, the fact that the spectrum is polluted with
high--frequency noise from the numerical error is further evidence
that error cancelation is the cause of the seemingly good accuracy. With ELES,
we see a that the filtering controls the noise in the tail of the spectrum,
leaving the accurate network to provide the correct amount of dissipation to
match the mean flow statistics.

\section{Conclusions}\label{sec:4}
In this work, we begin by highlighting a consistency issue between online and
offline training that occurs in state-of-the-art neural network--based
methods for subgrid stress closure modeling in LES, clarifying observations
made in the literature that highlight the contradictory results stemming from
this issue. While this is alternatively explained as a failure to
account for LES discretization errors while learning the functional form, it 
is important to highlight that the networks are faced with a dataset shift,
which is what makes them uniquely sensitive to numerical errors in a way that
physics--based closures are not.

We then propose a resolution to this inconsistency, by reformulating the 
problem using explicitly--filtered LES, which allows us to separate the 
numerical and modeling errors, using explicit filtering of the nonlinear term
to control the former, while allowing the networks to account for the latter.
We show that although this requires the networks to learn a more complex 
functional form --- since they now have to account for both the subgrid and
the subfilter scales --- the models now perform consistently with expectations
from \textit{a priori} accuracy studies. 

While the objective of the present work is not to make detailed comparisons of
neural network--based subfilter models with physics--based models for ELES, we
note here that the performance of physics--based closure models degrades when the LES is
explicitly filtered. The reason for this is twofold: firstly, as mentioned
earlier, ELES subfilter models need to account for more scales than the 
equivalent ILES subgrid models; secondly, some of the inadequacy of subgrid 
models is masked by numerical dissipation; thus, when this dissipation is 
controlled by explicit filtering, the ``true" performance of the model is 
observed. For instance, Gullbrand and Chow \cite{gull} found that the 
performance of the 
dynamic Smagorinsky model in a channel flow degraded significantly when the 
LES was explicitly filtered relative to when an implicit grid filter was used.

In theory, the approach presented here allows for a neural network--based 
closure model that has 
less solver--dependence, which is a key step in developing portable models that
do not require expensive offline re-training for use in different solvers.

\section*{Acknowledgments}
The authors gratefully acknowledge discussions with Sanjeeb Bose, Adrian Lozano-Dur\'{a}n
and Stefan Domino.

\section*{Declaration of competing interests}
The authors declare that they have no known competing financial interests or 
personal relationships that could have
appeared to influence the work reported in this paper.

\section*{Data availability}
The data used for training the neural networks 
as well as that used for generating the plots in this manuscript will be made available upon request.

\section*{Appendix A: Grid--independence of single--point SFS models for ELES}\label{app:a}
\begin{figure}%
    \centering
    \includegraphics[width=0.5\textwidth]{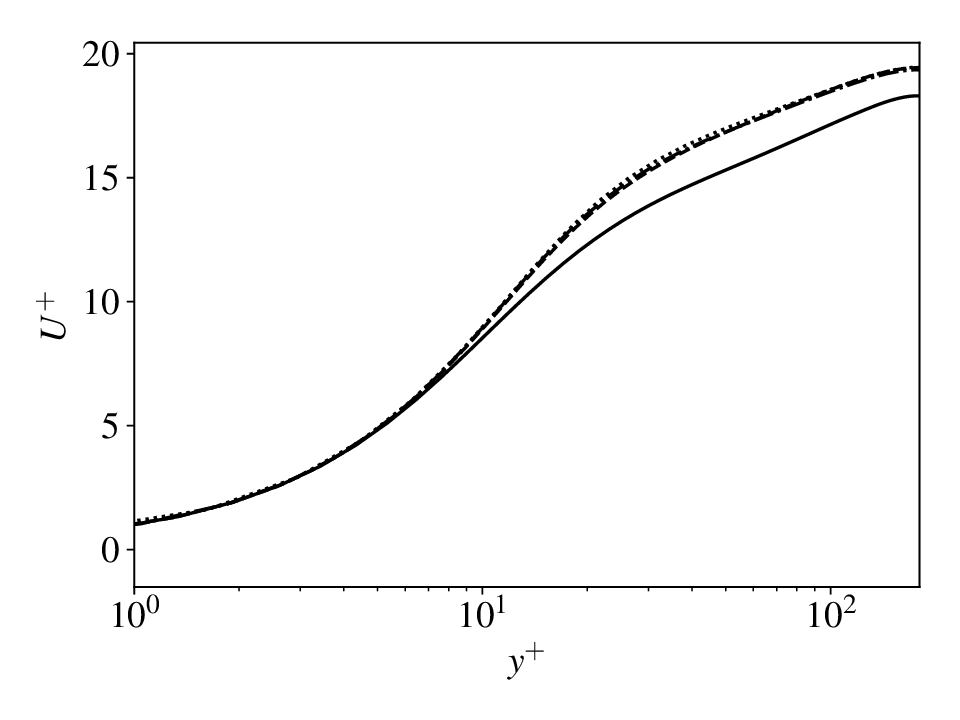} 
    \caption{Averaged streamwise velocity ($U^{+}$) profiles for the single--point
        ELES subfilter model. Solid line: DNS;
        dotted line: $48^3$ grid; dot--dashed line: $72^3$ grid; dashed line: 
        $96^3$ grid.}
    \label{fig:66}%
\end{figure}

In explicitly--filtered LES, the neural network learns a functional form that
relates filtered and residual quantities which is dependent only on the choice
of filter. This means that absent significant interference from the numerical
error in the low filter--to--width ratio simulations, the network should 
be performant on any grid, provided the same physical scales are filtered in
the same way. This is simplest to test using the low--accuracy single--point 
model, since the non--local 19--point model implicitly includes an additional
length scale that introduces grid--dependence (the average distance between 
each pair of the 19 nodes). Figure \ref{fig:66} shows the streamwise velocity
profiles for ELES using the single--point model on the three grids under 
consideration, and confirms this expection of grid--independence. This means
that while for the sake of accuracy the 19--point model is preferable, there
is a benefit to be had from a local model, as the present ELES formulation
avoids the need for training on several grids, an approach that is taken with
neural network--based models for ILES \cite{park}.
The other significant implication of the above result is that we can now use the 
$48^3$ grid for the 
neural network--based subfilter model, which significantly saves computational
cost, and makes the calculations as affordable as the ILES in terms of degrees
of freedom.

\section*{Appendix B: Spectra of ELES with and without subfilter models}
\begin{figure}%
    \centering
    \includegraphics[width=0.65\textwidth]{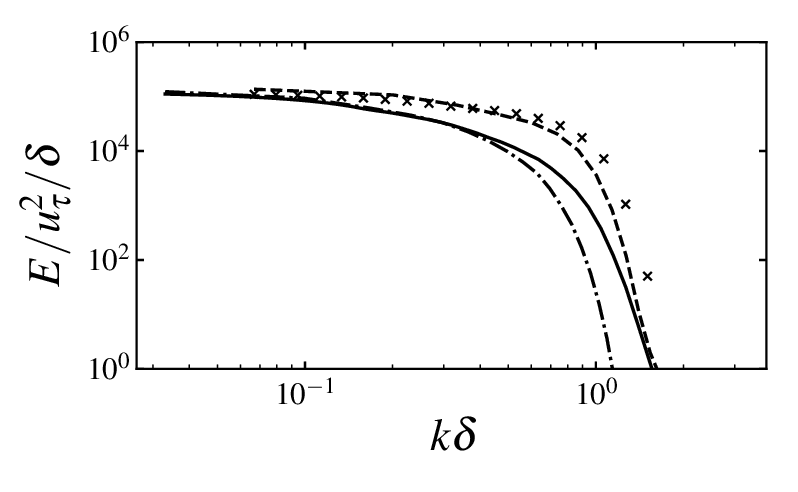} 
            \caption{Streamwise velocity spectrum at $y^+ \approx 180$, 
                non-dimensionalized by friction velocity ($u_{\tau}$) and
                channel half-height ($\delta$). Solid line: DNS; dot-dashed 
            lines: DNS filtered at $4\Delta$; dashed line: ELES with 19--pt model; crosses:
            ELES with no subfilter model; solid gray line: $-1$ power law.}
    \label{fig:63}%
\end{figure}

Figure \ref{fig:63} shows the streamwise velocity spectrum at the channel 
midplane, with a view to comparing the spectra obtained when running ELES with
and without a subfilter model. The 19--point ELES SFS model is shown. We see
that in the case with no subfilter model, there is an accumulation of energy
at the smaller scales, at a non-dimensional wavenumber of approximately unity.
The neural network--based subfilter model has the effect of drawing energy 
appropriately from the smallest resolved scales, which is consistent with the
role of a subfilter model.

\bibliographystyle{plain} 
\bibliography{refs}

\begin{thebibliography}{10}

\bibitem{akh}
R.~Akhavan, A.~Ansari, S.~Kang, and N.~Mangiavacchi.
\newblock Subgrid-scale interactions in a numerically simulated planar
  turbulent jet and implications for modelling.
\newblock {\em Journal of Fluid Mechanics}, 408:83–120, 2000.

\bibitem{stag}
H.~Jane Bae and Adrián Lozano-Durán.
\newblock Effect of wall boundary conditions on a wall-modeled large-eddy
  simulation in a finite-difference framework.
\newblock {\em Fluids}, 6(3), 2021.

\bibitem{mb}
M.~Benjamin, S.~Domino, and G.~Iaccarino.
\newblock Challenges in the use of neural networks for large-eddy simulations.
\newblock {\em CTR Annual Research Briefs}, pages 123--135, 2022.

\bibitem{grini}
J.P. Boris, F.F. Grinstein, E.S. Oran, and R.L. Kolbe.
\newblock New insights into large eddy simulation.
\newblock {\em Fluid Dynamics Research}, 10(4):199--228, 1992.

\bibitem{bosecomm}
Sanjeeb Bose.
\newblock Private communication, 2023.

\bibitem{bose}
Sanjeeb~T Bose, Parviz Moin, and Donghyun You.
\newblock Grid-independent large-eddy simulation using explicit filtering.
\newblock {\em Physics of Fluids}, 22(10):105103, 2010.

\bibitem{brunton}
Steven~L Brunton, Bernd~R Noack, and Petros Koumoutsakos.
\newblock Machine learning for fluid mechanics.
\newblock {\em Annu. Rev. Fluid Mech.}, 52:477--508, 2020.

\bibitem{chow}
Fotini~K Chow and Parviz Moin.
\newblock Explicit filtering for large eddy simulation on unstructured grids.
\newblock {\em Center for Turbulence Research Annual Research Briefs}, pages
  3--14, 2005.

\bibitem{chowmoin}
Fotini~Katopodes Chow and Parviz Moin.
\newblock A further study of numerical errors in large-eddy simulations.
\newblock {\em Journal of Computational Physics}, 184(2):366--380, 2003.

\bibitem{ferz}
Robert~A. Clark, Joel~H. Ferziger, and W.~C. Reynolds.
\newblock Evaluation of subgrid-scale models using an accurately simulated
  turbulent flow.
\newblock {\em Journal of Fluid Mechanics}, 91(1):1–16, 1979.

\bibitem{destef}
Giuliano De~Stefano, Oleg~V Vasilyev, and David~E Goldstein.
\newblock Effects of filter type and grid resolution on explicitly filtered
  large-eddy simulations.
\newblock {\em Flow, Turbulence and Combustion}, 74(3):261--284, 2005.

\bibitem{xiao}
Karthik Duraisamy, Gianluca Iaccarino, and Heng Xiao.
\newblock Turbulence modeling in the age of data.
\newblock {\em Annu. Rev. Fluid Mech.}, 51:357--377, 2019.

\bibitem{gamahara}
Masataka Gamahara and Yuji Hattori.
\newblock Searching for turbulence models by artificial neural network.
\newblock {\em Phys. Rev. Fluids}, 2, 2017.

\bibitem{dynamic}
Massimo Germano, Ugo Piomelli, Parviz Moin, and William~H. Cabot.
\newblock A dynamic subgrid‐scale eddy viscosity model.
\newblock {\em Phys. Fluids A: Fluid Dynamics}, 3(7):1760--1765, 1991.

\bibitem{ghosal}
Sandip Ghosal.
\newblock An analysis of numerical errors in large-eddy simulations of
  turbulence.
\newblock {\em Journal of Computational Physics}, 125(1):187--206, 1996.

\bibitem{kgoc}
Konrad~A Goc, Oriol Lehmkuhl, George~Ilhwan Park, Sanjeeb~T Bose, and Parviz
  Moin.
\newblock Large eddy simulation of aircraft at affordable cost: a milestone in
  computational fluid dynamics.
\newblock {\em Flow}, 1:E14, 2021.

\bibitem{subel}
Yifei Guan, Ashesh Chattopadhyay, Adam Subel, and Pedram Hassanzadeh.
\newblock Stable a posteriori les of 2d turbulence using convolutional neural
  networks: Backscattering analysis and generalization to higher $re$ via
  transfer learning.
\newblock {\em Journal of Computational Physics}, 458:111090, 2022.

\bibitem{gull}
Jessica Gullbrand and Fotini~Katopodes Chow.
\newblock The effect of numerical errors and turbulence models in large-eddy
  simulations of channel flow, with and without explicit filtering.
\newblock {\em Journal of Fluid Mechanics}, 495:323–341, 2003.

\bibitem{dropout}
Geoffrey~E Hinton, Nitish Srivastava, Alex Krizhevsky, Ilya Sutskever, and
  Ruslan Salakhutdinov.
\newblock Improving neural networks by preventing co-adaptation of feature
  detectors.
\newblock {\em arXiv preprint arXiv:1207.0580}, 2012.

\bibitem{gorle}
Yunjae Hwang and Catherine Gorlé.
\newblock Large-eddy simulations to define building-specific similarity
  relationships for natural ventilation flow rates.
\newblock {\em Flow}, 3:E10, 2023.

\bibitem{kang}
Myeongseok Kang, Youngmin Jeon, and Donghyun You.
\newblock Neural-network-based mixed subgrid-scale model for turbulent flow.
\newblock {\em arXiv preprint arXiv:2205.10181}, 2022.

\bibitem{kochov}
Dmitrii Kochkov, Jamie~A. Smith, Ayya Alieva, Qing Wang, Michael~P. Brenner,
  and Stephan Hoyer.
\newblock Machine learning–accelerated computational fluid dynamics.
\newblock {\em Proceedings of the National Academy of Sciences},
  118(21):e2101784118, 2021.

\bibitem{k41}
Andrei~N. Kolmogorov.
\newblock The local structure of turbulence in incompressible viscous fluid for
  very large reynolds number.
\newblock {\em Dokl. Akad. Nauk. SSSR}, 30:301--303, 1941.

\bibitem{lenny}
A.~Leonard.
\newblock Energy cascade in large-eddy simulations of turbulent fluid flows.
\newblock In F.N. Frenkiel and R.E. Munn, editors, {\em Turbulent Diffusion in
  Environmental Pollution}, volume~18 of {\em Advances in Geophysics}, pages
  237--248. Elsevier, 1975.

\bibitem{liu}
Bo~Liu, Huiyang Yu, Haibo Huang, Nansheng Liu, and Xiyun Lu.
\newblock Investigation of nonlocal data-driven methods for subgrid-scale
  stress modeling in large eddy simulation.
\newblock {\em {AIP} Advances}, 12(6):065129, 2022.

\bibitem{lund}
T.S. Lund.
\newblock The use of explicit filters in large eddy simulation.
\newblock {\em Comput. Math. with Appl.}, 46(4):603--616, 2003.
\newblock Turbulence Modelling and Simulation.

\bibitem{mars}
Alison~L Marsden.
\newblock Construction of commutative filters for les on unstructured meshes.
\newblock {\em Journal of computational physics}, 175(2), 2002.

\bibitem{osan}
R.~Maulik, O.~San, A.~Rasheed, and P.~Vedula.
\newblock Subgrid modelling for two-dimensional turbulence using neural
  networks.
\newblock {\em Journal of Fluid Mechanics}, 858:122–144, 2019.

\bibitem{mene}
Charles Meneveau and Joseph Katz.
\newblock Scale-invariance and turbulence models for large-eddy simulation.
\newblock {\em Annual Review of Fluid Mechanics}, 32(1):1--32, 2000.

\bibitem{moin}
Parviz Moin and John Kim.
\newblock Numerical investigation of turbulent channel flow.
\newblock {\em Journal of Fluid Mechanics}, 118:341–377, 1982.

\bibitem{moser}
Robert Moser, John Kim, and Nagi Mansour.
\newblock Direct numerical simulation of turbulent channel flow up to re=590.
\newblock {\em Phys. Fluids}, 11:943--945, 04 1999.

\bibitem{park}
Jonghwan Park and Haecheon Choi.
\newblock Toward neural-network-based large eddy simulation: application to
  turbulent channel flow.
\newblock {\em J. Fluid Mech.}, 914:A16, 2021.

\bibitem{prakash}
Aviral Prakash, Kenneth~E. Jansen, and John~A. Evans.
\newblock Invariant data-driven subgrid stress modeling in the strain-rate
  eigenframe for large eddy simulation.
\newblock {\em Computer Methods in Applied Mechanics and Engineering},
  399:115457, 2022.

\bibitem{rogallo}
Robert~S. Rogallo and Parviz Moin.
\newblock Numerical simulation of turbulent flows.
\newblock {\em Annual Review of Fluid Mechanics}, 16:99--137, 1984.

\bibitem{sreeni}
K.R. Sreenivasan.
\newblock On the universality of the kolmogorov constant.
\newblock {\em Physics of Fluids}, 7(11):2778--2784, 1995.

\bibitem{gmd}
R.~Stoffer, C.~M. van Leeuwen, D.~Podareanu, V.~Codreanu, M.~A. Veerman,
  M.~Janssens, O.~K. Hartogensis, and C.~C. van Heerwaarden.
\newblock Development of a large-eddy simulation subgrid model based on
  artificial neural networks: a case study of turbulent channel flow.
\newblock {\em Geosci. Model Dev.}, 14(6):3769--3788, 2021.

\bibitem{stolz}
S~Stolz, NA~Adams, and L~Kleiser.
\newblock An approximate deconvolution model for large-eddy simulation with
  application to incompressible wall-bounded flows.
\newblock {\em Physics of Fluids}, 13(4):997--1015, 2001.

\bibitem{toosi}
Amirhosein Toosi, Andrea~G Bottino, Babak Saboury, Eliot Siegel, and Arman
  Rahmim.
\newblock A brief history of ai: how to prevent another winter (a critical
  review).
\newblock {\em PET clinics}, 16(4):449--469, 2021.

\bibitem{thuerey}
Kiwon Um, Robert Brand, Yun~(Raymond) Fei, Philipp Holl, and Nils Thuerey.
\newblock Solver-in-the-loop: Learning from differentiable physics to interact
  with iterative pde-solvers.
\newblock In {\em Proceedings of the 34th International Conference on Neural
  Information Processing Systems}, Red Hook, NY, USA, 2020. Curran Associates
  Inc.

\bibitem{vasiada}
Oleg~V Vasilyev.
\newblock An adaptive version of the stretched-vortex subgrid-scale model for
  les on unstructured grids.
\newblock {\em Journal of Computational Physics}, 173(1):139--164, 2001.

\bibitem{vasy}
Oleg~V. Vasilyev, Thomas~S. Lund, and Parviz Moin.
\newblock A general class of commutative filters for les in complex geometries.
\newblock {\em J. Comput. Phys.}, 146(1):82--104, 1998.

\bibitem{xiec}
Chenyue Xie, Jianchun Wang, and Weinan E.
\newblock Modeling subgrid-scale forces by spatial artificial neural networks
  in large eddy simulation of turbulence.
\newblock {\em Phys. Rev. Fluids}, 5:054606, May 2020.

\bibitem{xie}
Chenyue Xie, Jianchun Wang, Hui Li, Minping Wan, and Shiyi Chen.
\newblock Artificial neural network mixed model for large eddy simulation of
  compressible isotropic turbulence.
\newblock {\em Phys. Fluids}, 31(8), 2019.

\bibitem{zhou}
Zhideng Zhou, Guowei He, Shizhao Wang, and Guodong Jin.
\newblock Subgrid-scale model for large-eddy simulation of isotropic turbulent
  flows using an artificial neural network.
\newblock {\em Comput. Fluids}, 195, 2019.

\end{thebibliography}

\end{document}